\documentclass{article}
\usepackage[latin1]{inputenc}
\usepackage[english]{babel}
\usepackage{dsfont}
\usepackage[cyr]{aeguill}
\usepackage{amsmath,amssymb}
\usepackage{amsthm}
\usepackage{eurosym}
\usepackage{multicol}
\usepackage{color}
\usepackage{fancyhdr}
\usepackage[pdftex]{graphicx}
\usepackage[pdftex]{hyperref}
\usepackage{verbatim}
\usepackage{setspace}
\usepackage{stmaryrd}

\newcommand{\proba}{\mathbb{P}} 
\newcommand{\indic}{\mathds{1}} 
\newcommand{\E}{\mathbb{E}} 
\DeclareMathOperator{\Cov}{Cov} 
\DeclareMathOperator{\Var}{Var} 
\DeclareMathOperator{\corr}{corr} 
\DeclareMathOperator{\sgn}{sgn} 
\DeclareMathOperator{\median}{med}
\DeclareMathOperator{\argmin}{argmin}

\theoremstyle{plain}
\newtheorem{thm}{Theorem}
\newtheorem{pro}{Proposition}

\newtheorem{lem}{Lemma}

\newtheorem{assum}{Assumption}
\providecommand{\keywords}[1]{\textbf{\textit{Keywords --}} #1}

\title{Estimation of bid-ask spreads \\ in the presence of serial dependence}

\author{Xavier Brouty$^{\text{a}}$, Matthieu Garcin$^{\text{b,c,}}$\thanks{Corresponding author: matthieu.garcin@m4x.org. \newline $^{\text{a}}$ ESILV, 92916 Paris La Défense, France. \newline $^{\text{b}}$ Léonard de Vinci Pôle Universitaire, Research center, 92916 Paris La Défense, France. \newline $^{\text{c}}$ This research benefited from the support of the Chaire ``Deep Finance Statistics'' between QRT, Ecole Polytechnique and its foundation.
\newline Acknowledgments: The authors thank Charles-Albert Lehalle as well as the participants of the 2024 Mathematical and statistical methods for actuarial sciences and finance conference, Le Havre, 2024 Econophysics colloquium, Vienna, 2024 Quantitative methods in finance conference, Sydney, for useful discussions and comments.}, Hugo Roccaro$^{\text{a}}$} 

\date{\today}

\begin{document}

\maketitle

\begin{abstract}
Starting from a basic model in which the dynamic of the transaction prices is a geometric Brownian motion disrupted by a microstructure white noise, corresponding to the random alternation of bids and asks, we propose moment-based estimators along with their statistical properties. We then make the model more realistic by considering serial dependence: we assume a geometric fractional Brownian motion for the price, then an Ornstein-Uhlenbeck process for the microstructure noise. In these two cases of serial dependence, we propose again consistent and asymptotically normal estimators. All our estimators are compared on simulated and real data with existing approaches, such as Roll, Corwin-Schultz, Abdi-Ranaldo, or Ardia-Guidotti-Kroencke estimators.
\end{abstract}




\section{Introduction}

When a simple model is introduced to solve a financial problem, incorporating frictions, namely liquidity risk, is one of its most natural extensions. For example, one can think about the design of optimal trading strategies mitigating market impacts~\cite{AC}, the definition of risk measures that include liquidity risk~\cite{BDS,AnB}, as well as the extensions of the Black-Scholes model for a non-zero bid-ask spread of the underlying, leading to specific arbitrage-free option prices~\cite{GP17,GG} and replication techniques~\cite{Leland,AP}. For applying all these methods, the accurate knowledge of the value of the bid-ask spread is of paramount importance.

Thanks to limit order books, the bid-ask spread is publicly displayed for major stock markets or is obtained easily as soon as one is able to match high-frequency trades and quote data~\cite{HJ}. Unfortunately, for many other tradable assets, bid-ask spreads are only latent variables. Investors thus need statistical methods to estimate the bid-ask spread in many situations: in over-the-counter transactions, illiquid markets such as those of corporate bonds~\cite{GP}, or dark pools~\cite{CM}. Moreover, even in a market with a public limit order book, low-frequency investors may be interested in an aggregated bid-ask spread, for example at a one-day time scale, more than in the tick-by-tick dynamic of limit order books. This kind of investor thus definitely needs bid-ask spread estimators based on time series of prices

The literature on the estimation of bid-ask spreads from a time series of displayed prices started with Roll's estimator, which is based on the empirical covariance of successive price increments~\cite{Roll}. The observable price is considered to be the sum of the mid price, that is the average between the bid and the ask, and a microstructure noise corresponding to a discrete variable equal to $-S/2$ or $S/2$, where $S$ is the bid-ask spread. Many alternatives to Roll's estimator have been proposed since, exploring approaches based on high-low ranges instead of cross moments~\cite{CS} and refinements related to overnight price movements~\cite{AR} or infrequent trading~\cite{AGK}. In this last case, when two consecutive observations of the time series of prices are the same because of an absence of intermediate trades, a spurious correlation of the microstructure noise appears. A straightforward correction of the spread estimator makes it possible to take this simple situation into account~\cite{AGK}. But a more general serial dependence introduces a bias for all the existing estimators cited above.

The presence of serial dependence in price returns, although rarely strong, has been documented in several empirical studies: in the literature of factor models~\cite{FF,HS}, in statistical studies underlining the only existence of such a serial dependence for short time scales of less than 20 minutes~\cite{Cont}, which can stem from the mean-reversion of the order book imbalance~\cite{LN}, or in the analysis of Hurst exponents, which shows that, if this idea is controversial for daily data of major stock indices~\cite{STG,TAM}, it is significant at least at short time scales for FX rates~\cite{Garcin17,Garcin20}. All these approaches designed to quantify serial dependence are based on linear models. Nonlinear models and model-free approaches, stemming for example from copula~\cite{RPS,Smith} or from information theory~\cite{AE,BG23,BG24}, tend to confirm the presence of serial dependence for many asset classes, depending on the time scale and the period of the study. On the other hand, the serial dependence of the sign of the trades seems to be a well-established fact, observable at high frequencies~\cite{AMZ,JLZ,BBD}.

It is thus difficult to reject the presence of any serial dependence in price returns or in microstructure noise. In this context, the introduction of estimators of bid-ask spreads that are robust to the existence of this serial dependence is an important issue. This is the purpose of this article. We propose several estimators. Some of them are appropriate in the absence of serial dependence. Others are built to take into account the presence of serial dependence either of price returns, or of trades, or of both. The asymptotic properties of these estimators are presented. A study on simulated data shows their relevance compared to classical bid-ask spread estimators. All our estimators are quite simple, insofar as they only rely on moments of close price increments, unlike classical methods, which use cross moments or high-frequency data summarized in high-low ranges.

To keep the framework as simple as possible, we have made the choice to depict the serial dependence of price returns thanks to a model with linear dependence and a low number of parameters, namely the fractional Brownian motion (fBm). This model of log-prices is widespread in finance and has been much studied, in particular in the context of statistical arbitrage~\cite{RTS,GNR,GMR,Garcin22}. The model for serially dependent trades is derived from an Ornstein-Uhlenbeck process, following the classical approach~\cite{BBD}.

In addition to the investigation of estimators of bid-ask spreads in the presence of serial dependence, we have also obtained two side results. First, a new estimator of Hurst exponent in the presence of microstructure noise, will enrich the literature on the estimation of fractional processes. Second, we introduce a decomposition of correlated Rademacher variables using independent variables following a more general two-point distribution.

The rest of the paper is organized as follows. In Section~\ref{sec:estim}, we introduce the various market models with or without serial dependence. For each of these models, we propose appropriate estimators for the bid-ask spread along with their properties. A simulation study and an application to a real financial dataset are presented in Section~\ref{sec:simul}. Section~\ref{sec:conclusion} concludes.

\section{Market models and spread estimation}\label{sec:estim}

This section contains all the theoretical results related to the estimators. We present the financial framework with various models and their corresponding estimators.

\subsection{General statement of the problem}

We introduce a market model, in which the mid price is not observed and evolves according to a stochastic process $P_t^{\star}$, whereas the observed price $P_t$, since it corresponds to the price of transactions, deviates from $P_t^{\star}$ by half the bid-ask spread, $S$. Consistently with most of the literature on the subject~\cite{Roll,CS,AR,AGK}, we assume that the spread is constant over the estimation period, which is typically one day, and that buys and sells are equally likely. We thus have the following dynamic of prices, for $t\geq 0$:
\begin{equation} \label{eq:price}
P_t = P_t^{\star}(1+S(X_t-0.5)),
\end{equation}
where $X_t \sim \mathcal{B}(1/2)$ is the trade direction. Considering the logarithmic price $p_t$, equation~\eqref{eq:price} becomes:
\begin{equation} \label{eq:logprice}
p_t = p_t^{\star}+\eta_t,
\end{equation}
where $\eta_t = S(X_t-0.5)$ describes the microstructure noise and the process $p_t^{\star}$ is the logarithmic mid price. Equation~\eqref{eq:logprice} is a common approximation in which one neglects the rest of the Taylor expansion of the logarithm of equation~\eqref{eq:price}. The smaller the spread, the better the approximation. We also note that for short time horizons, the Bachelier model is more relevant than the geometric Brownian motion approach, making equation~\eqref{eq:logprice} appropriate for describing prices instead of log-prices~\cite[Section 2.1.1]{BBD}.

The purpose of the paper is the estimation of $S$ based on the time series of observed log-prices. In what follows, we mainly introduce four estimators. Each of them corresponds to a particular set of assumptions on the dynamic defined in equation~\eqref{eq:logprice}, regarding the presence or not of serial dependence. So we successively examine the case where non-overlapping increments of $p_t^{\star}$ are independent of each other, and the case where they are correlated, modelled by a log-price following an fBm. In addition to this first source of serial dependence, we also investigate the case where the trades are autocorrelated, that is where the microstructure noise $\eta_t$ is autocorrelated, with an exponentially decaying autocorrelation function, stemming for instance from a mean-reverting underlying process that follows the Ornstein-Uhlenbeck model. Apart from these two serial dependences, we always make the following assumption.

\begin{assum}\label{assum:indepProcesses}
The processes $p_t^{\star}$ and $\eta_t$ are independent of each other.
\end{assum}

Note that we work with estimators of $S^2$ instead of $S$. In some situations, this estimator may be negative. As it is usually the case in the literature about spread estimation, we can propose a transformation of this estimator intended to obtain a positive spread, such as $\widehat S=\max(0,\widehat{S^2})^{1/2}$, which is the method adopted in our simulation study in Section~\ref{sec:simul}, or $\widehat S=|\widehat{S^2}|^{1/2}$~\cite{GHT,AGK}. Both these transformations are continuous but non-differentiable in zero. This means that the property of consistency of the estimator $\widehat{S^2}$ will also apply to the estimator $\widehat S$, thanks to the continuous mapping theorem. If we assume that $S\neq 0$, we can also extend the property of asymptotic normality from $\widehat{S^2}$ to $\widehat S$ thanks to the delta method, which only requires the differentiability of the above transformation around $S$~\cite[Theorem 3.1]{vanderVaart}.

The estimation of $S^2$ is based on observations of $p_t$ at discrete times. We assume we have $n$ observations sampled with a time step $\tau>0$: $p_{0}, p_{\tau},...,p_{(n-1)\tau}$. In what follows, $\tau$ is typically equal to one minute.

\subsection{Estimators for various dependence assumptions}\label{sec:estimators}

In what follows, we successively make several assumptions regarding the presence of serial dependence and we introduce a bid-ask spread estimator adapted to each model, along with its asymptotic properties.

\subsubsection{Standard zero-autocorrelation market model}\label{sec:zeroCorrel}

The first model we investigate is the simplest and corresponds to the following assumption. 

\begin{assum}\label{assum:zeroCorrel}
We assume independence both for the increments of $p_t^{\star}$, with a Gaussian specification, and for the microstructure noise:
\begin{itemize}
\item $p_t^{\star}$ is a Brownian motion with volatility parameter $\sigma>0$,
\item $\eta_s$ and $\eta_t$ are independent of each other as soon as $s\neq t$.
\end{itemize}
\end{assum}

The aim of our work is to estimate the parameter $S$ in equation~\eqref{eq:logprice}, that is the bid-ask spread. We propose to address this estimation problem with a moment method. In particular, using the above assumptions of serial independence and Gaussian increments of $p_t^{\star}$, as well as the independence between $p_t^{\star}$ and $\eta_t$ stated in Assumption~\ref{assum:indepProcesses}, we get the theoretical variance of the price log-returns, whatever $\tau\geq 0$:
\begin{equation}\label{eq:varianceIncrPrix}
V(L\tau)=\Var(p_{t+L\tau}-p_t)=L\tau\sigma^2 + \frac{S^2}{2},
\end{equation}
where $L\in\llbracket 1,n\rrbracket$. Both the parameters $\sigma^2$ and $S^2$ appear in this expression, so that an estimator of $S^2$ alone would require combining the moments at two time scales $L$ and $L'$ to get rid of $\sigma^2$:
\begin{equation}\label{eq:S_vs_moments1}
L'V(L\tau)-LV(L'\tau)=\frac{S^2}{2}(L'-L).
\end{equation}

In order to estimate $\Var(p_{t+L\tau}-p_t)$, we use the empirical variance with zero mean and either overlapping increments, $\widehat V_1(n,L)$, or non-overlapping increments, $\widehat V_2(n,L)$, or strictly disjoint increments, $\widehat V_3(n,L)$:
$$\left\{\begin{array}{l}
\widehat V_1(n,L)=\frac{1}{k_1(n,L)}\sum_{i=0}^{k_1(n,L)-1}\left(p_{(i+L)\tau}-p_{i\tau}\right)^2 \\
\widehat V_2(n,L)=\frac{1}{k_2(n,L)}\sum_{i=0}^{k_2(n,L)-1}\left(p_{(i+1)L\tau}-p_{iL\tau}\right)^2 \\
\widehat V_3(n,L)=\frac{1}{k_3(n,L)}\sum_{i=0}^{k_3(n,L)-1}\left(p_{(i+(i+1)L)\tau}-p_{(i+iL)\tau}\right)^2,
\end{array}\right.$$
where $k_1(n,L)=n-L$, $k_2(n,L)=\lfloor(n-1)/L\rfloor$, and $k_3(n,L)=\lfloor n/(L+1)\rfloor$. At first glance, these three estimators should have different statistical properties because they are based on the empirical variance of either dependent or independent increments. The dependence can stem from two trade directions which repeat themselves at two opposite bounds of successive increments, like in $\widehat V_2(n,L)$, or also from a common diffusion part in two overlapping increments, like in $\widehat V_1(n,L)$. In order to determine the moments of these three estimators, we thus need the covariance of two squared price increments
\begin{equation}\label{eq:covIncrSq}
K(u,\delta)=\E\left[(p_{u}-p_0)^2(p_{2u-\delta}-p_{u-\delta})^2\right],
\end{equation}
for $u\geq 0$ and $\delta\leq u$, the case $\delta<0$ corresponding to strictly disjoint increments. This is the purpose of Proposition~\ref{pro:covIncrSq1}, whose proof is postponed in Appendix~\ref{sec:covIncrSq1}.

\begin{pro}\label{pro:covIncrSq1}
Under the market model introduced in equation~\eqref{eq:logprice} and Assumptions~\ref{assum:indepProcesses} and~\ref{assum:zeroCorrel}, the covariance $K(u,\delta)$ introduced in equation~\eqref{eq:covIncrSq}, for $u\geq 0$, is such that
$$K(u,\delta)=\left\{\begin{array}{ll}
V(u)^2 & \text{if } \delta\leq 0 \\
V(u)^2+2\sigma^4\delta^2 & \text{if } \delta\in[0,u) \\
V(u)^2 + 2\sigma^4u^2 + 2\sigma^2uS^2 + \frac{S^4}{4} & \text{if } \delta=u,
\end{array}\right.$$
where $V$ is defined in equation~\eqref{eq:varianceIncrPrix}.
\end{pro}

Using Proposition~\ref{pro:covIncrSq1}, one is then able to determine the first two moments of the three empirical variance estimators.

\begin{pro}\label{pro:momentV1}
Under the market model introduced in equation~\eqref{eq:logprice} and Assumptions~\ref{assum:indepProcesses} and~\ref{assum:zeroCorrel}, we have
$$\E\left[\widehat V_v(n,L)\right]=V(L\tau)$$
for all $v\in\{1,2,3\}$, where $V$ is defined in equation~\eqref{eq:varianceIncrPrix}. Moreover, the variance of the empirical variance estimator is $\Var\left[\widehat V_1(n,L)\right]=\sigma^2_{1,1}(L)/k_1(n,L) + \mathcal{O}\left(n^{-2}\right)$, where
$$\sigma^2_{1,1}(L)=\frac{2}{3}\sigma^4L(1+2L^2)\tau^2 + 2\sigma^2L\tau S^2 + \frac{S^4}{4},$$
and, if $v\in\{2,3\}$, $\Var\left[\widehat V_v(n,L)\right]=\sigma^2_{1,v}(L)/k_v(n,L)$, where
$$\sigma^2_{1,v}(L)=2\sigma^4L^2\tau^2 + 2\sigma^2L\tau S^2 + \frac{S^4}{4}.$$
\end{pro}

The proof of Proposition~\ref{pro:momentV1} is postponed in Appendix~\ref{sec:momentV1}. It mainly relies on a side property, detailed in Appendix~\ref{sec:lemmas}, regarding the decomposition of correlated Rademacher variables in independent variables following a more general two-point distribution.

We then use these empirical variances to define an estimator for $S^2$ in this first specification of the market model, for $L\neq L'$ and $v\in\{1,2,3\}$:
\begin{equation}\label{eq:S1}
\widehat{S^2_{1,v}}(n,L,L')=\frac{2}{L'-L}\left(L'\widehat V_v(n,L)-L\widehat V_v(n,L')\right).
\end{equation}
Theorem~\ref{thm:S1} provides some properties of this estimator of $S^2$.

\begin{thm}\label{thm:S1}
Under the market model introduced in equation~\eqref{eq:logprice} and Assumptions~\ref{assum:indepProcesses} and~\ref{assum:zeroCorrel}, the statistic $\widehat{S^2_{1,v}}(n,L,L')$ introduced in equation~\eqref{eq:S1}, for $v\in\{1,2,3\}$ and $L$ and $L'$ fixed integers such that $L\neq L'$, is an unbiased and consistent estimator of $S^2$. When $v=3$, it is a strong consistency. Moreover, 
\begin{equation}\label{eq:TCL1}
\sqrt{n}\left(\widehat{S^2_{1,v}}(n,L,L') - S^2\right)\overset{d}{\longrightarrow} \mathcal N\left(0,\gamma_{1,v}(L,L')\right),
\end{equation}
where $\overset{d}{\longrightarrow}$ stands for the convergence in distribution when $n\rightarrow \infty$, $\mathcal N(a,b)$ is the Gaussian distribution of mean $a$ and variance $b$, and
$$\gamma_{1,v}(L,L') = \frac{4}{(L'-L)^2} \left(L'^2\zeta_v(L)\sigma_{1,v}^2(L) + L^2\zeta_v(L')\sigma_{1,v}^2(L') - 2LL'\sqrt{\zeta_v(L)\zeta_v(L')}\sigma_{1,v}(L)\sigma_{1,v}(L')r_{1,v}(L,L')\right),$$
with $\sigma_{1,v}(L)$ defined as in Proposition~\ref{pro:momentV1}, $r_{1,v}(L,L')=\lim_{n\rightarrow\infty}\corr\left(\widehat V_v(n,L),\widehat V_v(n,L')\right)$, and $(\zeta_1(L),\zeta_2(L),\zeta_3(L))=(1,L,L+1)$. In particular, for all fixed $m\in\mathbb N\setminus\{0,1\}$, we have $\gamma_{1,3}(L,m(L+1)-1)=\Gamma(L,m(L+1)-1)$, where
$$\begin{array}{ccl}
\Gamma(L,L') & = & \frac{4(L'+1)}{(L'-L)^2} \left\{L'\left(\frac{L'(L+1)}{L'+1}-2L\right)\left(K(L\tau,L\tau)-V(L\tau)^2\right) + L^2\left(K(L'\tau,L'\tau)-V(L'\tau)^2\right) \right. \\
 & & \left.- 2L'L\left(-2\frac{L'-L}{L'+1}S^2V(L\tau)+\frac{3L'-4L-1}{L'+1}\frac{S^4}{4}\right)\right\}.
\end{array}$$
\end{thm}

The proof of Theorem~\ref{thm:S1} is postponed in Appendix~\ref{sec:S1}. The proof of the asymptotic normality relies on the application of the multivariate delta method and on the extension of the central limit theorem to $\alpha$-mixing random variables. The obtained expression, in equation~\eqref{eq:TCL1}, requires the knowledge of $r_{1,v}(L,L')$. The exact calculation of this correlation is very fastidious and many cases are to be taken into account if one wants to be exhaustive, but it does not contain big technical difficulties. We thus have decided instead to give a simple example in the theorem, corresponding to the case $v=3$ and $L'=m(L+1)-1$, which, in contrast to other choices of $v$, $L$, and $L'$, leads to a quite concise result. For a practical application of Theorem~\ref{thm:S1} with another choice of $v$, $L$, and $L'$, we suggest either to do the same kind of calculation as the one already detailed in the proof, or to replace the theoretical correlation $r_{1,v}(L,L')$ by its counterpart obtained by simulations, or even to evaluate by simulations the variance of $\widehat{S^2_{1,v}}(n,L,L')$. In this last approach, the importance of the theorem, beyond the knowledge of the exact parameters, is the statement of the Gaussian limit.

\begin{figure}[htbp]
	\centering
		\includegraphics[width=0.8\textwidth]{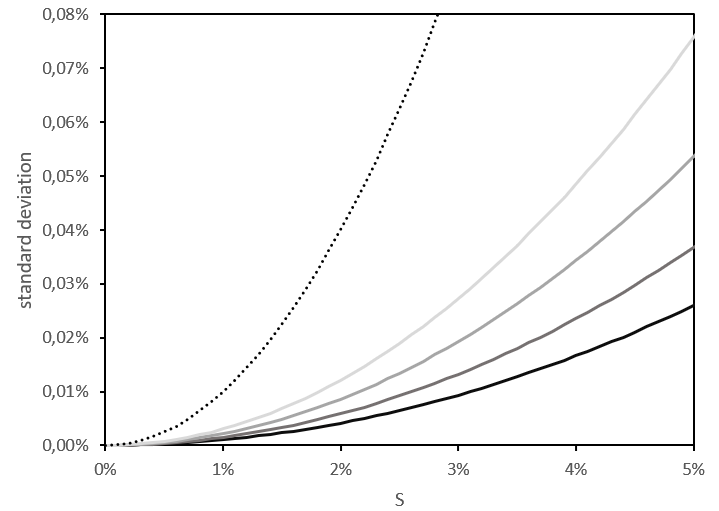} 
\begin{minipage}{0.7\textwidth}\caption{Asymptotic standard deviation of the estimator $\widehat{S^2_{1,3}}(n,1,3)$ as a function of the true spread $S$, for $\sigma=0.2$, $\tau=1/(260\times 510)$ corresponding to a one-minute time step, $n$ equal, from bottom to top, to 510 (observations during one trading day), 255, 120, 60 (observations in one hour). The dotted line corresponds to $S\mapsto S^2$.}
	\label{fig:StDevZeroCorrel}
\end{minipage}
\end{figure}

We represent in Figure~\ref{fig:StDevZeroCorrel} the standard deviation of the estimator of $S^2$ in the case where $m=2$, for which:
$$\Gamma(1,3) = 4 \left\{-\frac{3}{2}\left(K(\tau,\tau)-V(\tau)^2\right) + \left(K(3\tau,3\tau)-V(3\tau)^2\right) - 6\left(-S^2V(\tau)+\frac{S^4}{4}\right)\right\}.$$
We see in particular that the estimator is quite accurate whatever the spread, with a standard deviation far below the true $S^2$, even for only one hour of observations sampled every minute.

The estimator introduced in equation~\eqref{eq:S1} depends on the selection of the free parameters $L$ and $L'$. Low values for these two parameters lead to a higher number of price returns included in the estimator, so that one is inclined to choose $(L,L')=(1,2)$ to get a low variance for the estimator $\widehat{S^2_{1,v}}(n,L,L')$. But in practice, depending on the size and nature of the dataset, this estimator might be negative or have a too big variance. In this case, one may stack several estimators, using for example $\median\left\{\left.\widehat{S^2_{1,v}}(n,1,L)\right|L\in\llbracket 2,\mathcal L\rrbracket\right\}$, where $\median$ is the median and $\mathcal L\geq 2$ is an integer.

\subsubsection{Market model with autocorrelated price increments}\label{sec:fBm}

Among the models of serial dependence, the fBm is widespread in finance for depicting log-prices~\cite{BSV,Garcin22}. One of the reasons of this popularity stems from its parsimony: a single parameter, the Hurst exponent $H\in(0,1)$, is enough to describe serial dependence. The fBm has been introduced by Mandelbrot and Van Ness as a generalization of the Brownian motion~\cite{MvN} describing a non-zero correlation between non-overlapping increments. If $H>1/2$ (respectively $H<1/2$), the fBm $B^H_t$ is defined as the fractional integral (resp. derivative) of order $H-1/2$ (resp. $1/2-H$) of a Brownian motion $B_t$:
$$B^H_t=\frac{1}{\Gamma\left(H+\frac{1}{2}\right)}\int_{-\infty}^{\infty}{\left((t-s)_+^{H-1/2}-(-s)_+^{H-1/2}\right)dB_s}.$$
An equivalent definition of the fBm states that $B^H_t$ is the only Gaussian process such that $B^H_0=0$, $\E\left[B^H_t\right]=0$ for all $t$, and with the following variance of an increment between times $s$ and $t$:
$$\E\left[(B^H_t-B^H_s)^2\right]=|t-s|^{2H}.$$
The direct consequence of this last equation is that the covariance between two increments is
$$\E\left\{(B^H_t-B^H_s)(B^H_v-B^H_u)\right\}=\frac{1}{2}(|u-t|^{2H}+|v-s|^{2H}-|v-t|^{2H}-|u-s|^{2H}),$$
for all $s, t, u, v\in\mathbb R$. In the case of non-overlapping increments, this covariance is positive if $H>1/2$, negative if $H<1/2$, and equal to zero otherwise.

Several extensions of the fBm have been proposed, introducing stationarity~\cite{CKM,CV,FBA,Garcin19,Garcin22b,BG24}, replacing the Gaussian distribution by a more general one, in particular the stable distribution~\cite{ST94,WBM,AG,MP}, or considering a time-varying Hurst exponent following either a deterministic function~\cite{PLV,BRJ,Coeurjolly2005,ST06,BP,Garcin17} or a stochastic process~\cite{AT,BP11,Garcin20,LMS,AB}. These extensions offer the possibility to define more complex serial dependences. However, since price returns with serial dependence have not been studied yet in the literature devoted to the estimation of bid-ask spreads, we focus in this paper on the sole fBm.

In this problem of bid-ask spread estimation, we more precisely study the situation where log-prices and microstructure noise still follow equation~\eqref{eq:logprice} and Assumption~\ref{assum:indepProcesses}. In addition, we impose the following assumption.

\begin{assum}\label{assum:fBm}
We assume that the only serial dependence is for the increments of $p_t^{\star}$, with an fBm specification:
\begin{itemize}
\item $p_t^{\star}=\sigma B^H_t$, where $\sigma>0$ and $B^H_t$ is an fBm of Hurst exponent $H\in(0,1)$,
\item $\eta_s$ and $\eta_t$ are independent of each other as soon as $s\neq t$.
\end{itemize}
\end{assum}

Using Assumptions~\ref{assum:indepProcesses} and~\ref{assum:fBm}, the theoretical variance of the price log-returns becomes, whatever $\tau\geq 0$,
\begin{equation}\label{eq:varianceIncrPrix_fBm}
V(L\tau)=\Var(p_{t+L\tau}-p_t)=(L\tau)^{2H}\sigma^2 + \frac{S^2}{2},
\end{equation}
where $L\in\llbracket 1,n\rrbracket$. The parameters $\sigma^2$, $H$, and $S^2$ all appear in equation~\eqref{eq:varianceIncrPrix_fBm} but, following the same rationale as in Section~\ref{sec:zeroCorrel}, we can at least get rid of $\sigma^2$ by combining several moments. This leads to the following estimator of $S^2$, for $v\in\{1,2,3\}$:
\begin{equation}\label{eq:S2}
\widehat{S^2_{2,v}}(n,L,L',H)=\frac{2}{L'^{2H}-L^{2H}}\left(L'^{2H}\widehat V_v(n,L)-L^{2H}\widehat V_v(n,L')\right).
\end{equation}
This estimator depends on $H$. Forcing $H=1/2$ leads to the estimator of the zero-autocorrelation model, introduced in equation~\eqref{eq:S1}. But if the time series really follows Assumptions~\ref{assum:indepProcesses} and~\ref{assum:fBm}, the standard estimator $\widehat{S^2_{1,v}}(n,L,L')$ is biased. Indeed, with these assumptions, the variance of the log-price increment of duration $L\tau$ increases by the amount $((L\tau)^{2H}-L\tau)\sigma^2$ and the estimator of $S^2$ by the bias $2\sigma^2\tau^{2H}\frac{L'L^{2H}-LL'^{2H}}{L'-L}$, which is equal to $4\sigma^2\tau^{2H}(1-2^{2H-1})$ in the case where $L=1$ and $L'=2$, which is the one leading to the lowest bias. It is worth noting that Roll's estimator suffers from exactly the same bias. Indeed, starting from the expected value and variance of Roll's estimator~\cite{Harris}, one can easily show that the bias described above is $-4$ times the covariance of successive increments~\cite{Roll}. To conclude about this bias, our basic estimator of Section~\ref{sec:zeroCorrel} and Roll's estimator tend to overestimate (respectively underestimate) $S^2$ when $H<1/2$ (resp. $H>1/2$). The bias even explodes for very low values of $H$. This is our first observation that negative autocorrelation is the situation for which the need of new bid-ask spread estimators is the most striking. It will later be confirmed, in particular in the simulation study, in Section~\ref{sec:simul}.

Since the true Hurst exponent is unknown, we can first estimate $H$ then plug this estimator in equation~\eqref{eq:S2}. The estimation of $H$ can also rely on the mixing of various empirical variances. Indeed, using equation~\eqref{eq:varianceIncrPrix_fBm}, we note that
$$\left\{\begin{array}{ccl}
 V(2L\tau)-V(L\tau) & = & \sigma^2(L\tau)^{2H}(2^{2H}-1) \\
 V(4L\tau)-V(2L\tau) & = & \sigma^2(2L\tau)^{2H}(2^{2H}-1), 
\end{array}\right.$$
which naturally leads to the following estimator of $H$:
\begin{equation}\label{eq:estimH}
\widehat H_L=\frac{1}{2}\log_2\left|\frac{\widehat V_v(n,4L)-\widehat V_v(n,2L)}{\widehat V_v(n,2L)-\widehat V_v(n,L)}\right|.
\end{equation}

In order to determine the asymptotic properties of $\widehat{S^2_{2,v}}(n,L,L',\widehat H_{L''})$, we first need to study the function $K$ introduced in equation~\eqref{eq:covIncrSq} and the first two moments of the empirical variance estimators. This is the purpose of Propositions~\ref{pro:covIncrSq2} and~\ref{pro:momentV2}, whose proof is postponed in Appendix~\ref{sec:covIncrSq2} and~\ref{sec:momentV2}.

\begin{pro}\label{pro:covIncrSq2}
Under the market model introduced in equation~\eqref{eq:logprice} and Assumptions~\ref{assum:indepProcesses} and~\ref{assum:fBm}, the covariance $K(u,\delta)$ introduced in equation~\eqref{eq:covIncrSq}, for $u\geq 0$, is such that
$$K(u,\delta)=\left\{\begin{array}{ll}
V(u)^2 + 2c(\delta/u)^2u^{4H}\sigma^4 & \text{if } \delta\in(-\infty,0)\cup(0,u) \\
V(u)^2 + 2c(0)^2u^{4H}\sigma^4 -c(0)u^{2H}\sigma^2S^2 & \text{if } \delta=0 \\
V(u)^2+2u^{4H}\sigma^4+2u^{2H}\sigma^2S^2+S^4/4 & \text{if } \delta=u,
\end{array}\right.$$
where $V$ is defined in equation~\eqref{eq:varianceIncrPrix_fBm} and $c(x)=\frac{1}{2}\left(\left|2-x\right|^{2H} - 2\left|1-x\right|^{2H} + \left|x\right|^{2H}\right)$.
\end{pro}

\begin{pro}\label{pro:momentV2}
Under the market model introduced in equation~\eqref{eq:logprice} and Assumptions~\ref{assum:indepProcesses} and~\ref{assum:fBm}, we have
$$\E\left[\widehat V_v(n,L)\right]=V(L\tau)$$
for all $v\in\{1,2,3\}$, where $V$ is defined in equation~\eqref{eq:varianceIncrPrix_fBm}. Moreover, the variance of the empirical variance estimator is $\Var\left[\widehat V_1(n,L)\right]=\sigma^2_{2,1}(L)/k_1(n,L) + o(n^{-1})$ when $H\in(0,3/4)$, where
$$\sigma^2_{2,1}(L)=2(L\tau)^{4H}\sigma^4-(2L\tau)^{2H}\sigma^2S^2+\frac{S^4}{4}.$$
For $v\in\{2,3\}$, we have $\Var\left[\widehat V_2(n,L)\right]=\sigma^2_{2,2}(L)/k_2(n,L)+\xi_2(L,k_2(n,L))$,
where $\sigma^2_{2,2}(L)=\sigma^2_{2,1}(L)$ and
$$\xi_2(L,k)=\frac{4(L\tau)^{4H}\sigma^4}{k^2}\sum_{i=0}^{k-2} c(-i)^2(k-1-i)=\left\{\begin{array}{ll}
\mathcal O(k^{4H-4}) & \text{if } H\neq 3/4 \\
\mathcal O(\ln(k)/k) & \text{else,}
\end{array}\right.$$
and $\Var\left[\widehat V_3(n,L)\right]=\sigma^2_{2,3}(L)/k_3(n,L)+\xi_3(L,k_3(n,L))$,
where
$$\sigma^2_{2,3}(L)=2(L\tau)^{4H}\sigma^4+2(L\tau)^{2H}\sigma^2S^2+\frac{S^4}{4}$$
and
$$\xi_3(L,k)=\frac{4(L\tau)^{4H}\sigma^4}{k^2}\sum_{i=0}^{k-2} c\left(-i-\frac{i+1}{L}\right)^2(k-1-i)=\left\{\begin{array}{ll}
\mathcal O(k^{4H-4}) & \text{if } H\neq 3/4 \\
\mathcal O(\ln(k)/k) & \text{else.}
\end{array}\right.$$
\end{pro}

We can now use these properties on the variance estimator to establish the asymptotic behaviour of the bid-ask spread estimator, in the case where log-prices follow a fractional dynamics. It is the purpose of Theorem~\ref{thm:S2}, in which we get different results, depending on whether $H$ is known or estimated.

\begin{thm}\label{thm:S2}
Under the market model introduced in equation~\eqref{eq:logprice} and Assumptions~\ref{assum:indepProcesses} and~\ref{assum:fBm}, the statistics $\widehat{S^2_{2,v}}(n,L,L',H)$ and $\widehat{S^2_{2,v}}(n,L,L',\widehat H_{L''})$ introduced in equations~\eqref{eq:S2} and~\eqref{eq:estimH}, for $v\in\{1,2,3\}$ and $L$, $L'$, and $L''$ fixed integers such that $L\neq L'$, are consistent estimators of $S^2$. In addition, $\widehat{S^2_{2,v}}(n,L,L',H)$ is unbiased. Moreover, when $H\leq 3/4$,
$$\sqrt{n}\left(\widehat{S^2_{2,v}}(n,L,L',H) - S^2\right)\overset{d}{\longrightarrow} \mathcal N\left(0,\gamma_{2,v}(L,L',H)\right),$$
where
$$\begin{array}{ccl}
\gamma_{2,v}(L,L',H) & = & \frac{4}{(L'^{2H}-L^{2H})^2} \Bigl[L'^{4H}\zeta_v(L)\sigma_{2,v}^2(L) + L^{4H}\zeta_v(L')\sigma_{2,v}^2(L') \\
 & & - 2(LL')^{2H}\sqrt{\zeta_v(L)\zeta_v(L')}\sigma_{2,v}(L)\sigma_{2,v}(L')r_{2,v}(L,L')\Bigr].
\end{array}$$
with $\sigma_{2,v}(L)$ defined as in Proposition~\ref{pro:momentV2}, $r_{2,v}(L,L')=\lim_{n\rightarrow\infty}\corr\left(\widehat V_v(n,L),\widehat V_v(n,L')\right)$, and $(\zeta_1(L),\zeta_2(L),\zeta_3(L))$ defined as in Theorem~\ref{thm:S1}. We also have, for $H\leq 1/2$,
\begin{equation}\label{eq:TCL_fBm}
\sqrt{n}\left(\widehat{S^2_{2,v}}(n,1,2,\widehat H_1) - S^2\right)\overset{d}{\longrightarrow} \mathcal N\left(0,\frac{4}{\left(V(4\tau)-2V(2\tau)+V(\tau)\right)^4}\mathbf W^T\Sigma\mathbf W\right),
\end{equation}
where, for $i,j\in\llbracket 1,3\rrbracket$, $\Sigma_{ij}=\sqrt{\zeta_v(2^{i-1})\zeta_v(2^{j-1})}\sigma_{2,v}(2^{i-1})\sigma_{2,v}(2^{j-1})r_{2,v}(2^{i-1},2^{j-1})$ and
$$\mathbf W = \left(\begin{array}{c}
\left(V(4\tau)-V(2\tau)\right)^2 \\
2V(2\tau)\left(V(2\tau)-V(4\tau)-V(\tau)\right)+2V(4\tau)V(\tau) \\
\left(V(\tau)-V(2\tau)\right)^2
\end{array}\right).$$
\end{thm}

The proof of Theorem~\ref{thm:S2} is postponed in Appendix~\ref{sec:S2}. The main difference with the proof of Theorem~\ref{thm:S1} is that serial dependence prevents us from using the central limit theorem. Fortunately, fBm's inference is a well-known subject and asymptotic distributions are established for $H\leq 3/4$~\cite{Coeurjolly2001}. We thus use a central limit theorem for dependent variables, which is only valid if two distant observations of the process are asymptotically independent. This property does not hold for the fBm, as underlined in the pioneering literature on this model~\cite[Section 5.2]{MvN}, but considering discrete variations of this process, possibly variations of order higher than 1, makes it possible to restrict this dependence and thus to build an asymptotic theory~\cite{PV,IL,KW,Coeurjolly2001}. We show in the proof that adding to the fBm a microstructure noise following Assumptions~\ref{assum:indepProcesses} and~\ref{assum:fBm} does not prevent the empirical variances to converge toward Gaussian variables, but their parameters obviously depend on this microstructure noise.

The asymptotic distribution provided in Theorem~\ref{thm:S2} for $\widehat{S^2_{2,v}}(n,L,L',H)$, that is when $H$ is known, is very general. This contrasts with the case of $\widehat{S^2_{2,v}}(n,1,2,\widehat H_1)$, that is when $H$ is to be estimated. Indeed, in this case, for conciseness, we have focussed on a particular choice for $(L,L',L'')$, namely $(1,2,1)$. In addition, though $\widehat{S^2_{2,v}}(n,L,L',H)$ is unbiased, it is not the case for $\widehat{S^2_{2,v}}(n,L,L',\widehat H_{L''})$, which is only asymptotically unbiased.

\begin{figure}[htbp]
	\centering
		\includegraphics[width=0.45\textwidth]{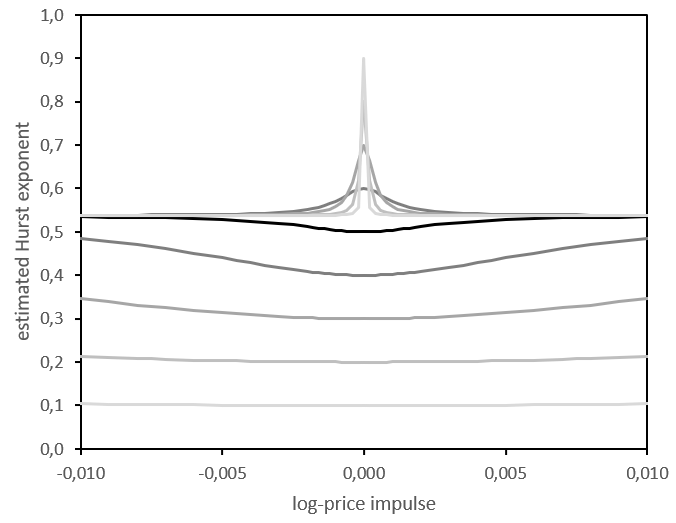} 
		\includegraphics[width=0.45\textwidth]{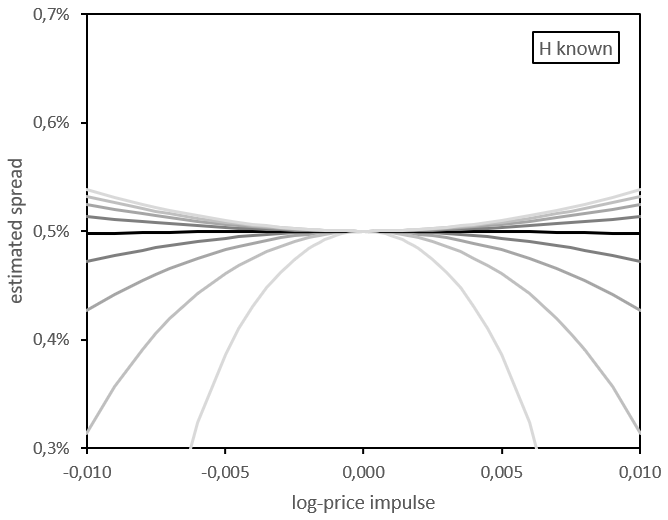} \\
		\includegraphics[width=0.45\textwidth]{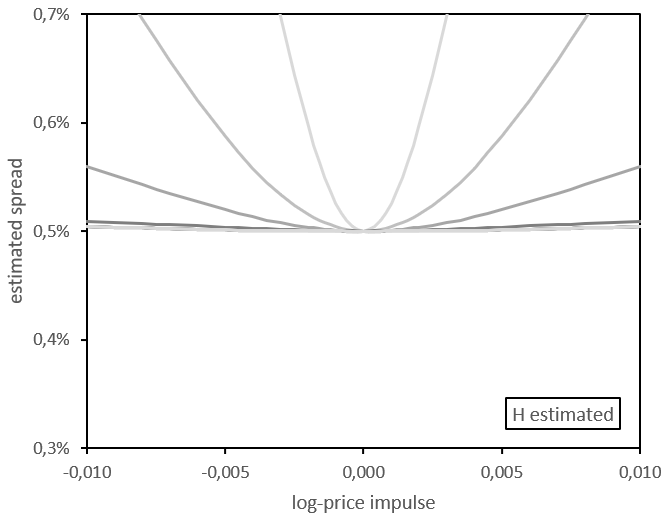} 
		\includegraphics[width=0.45\textwidth]{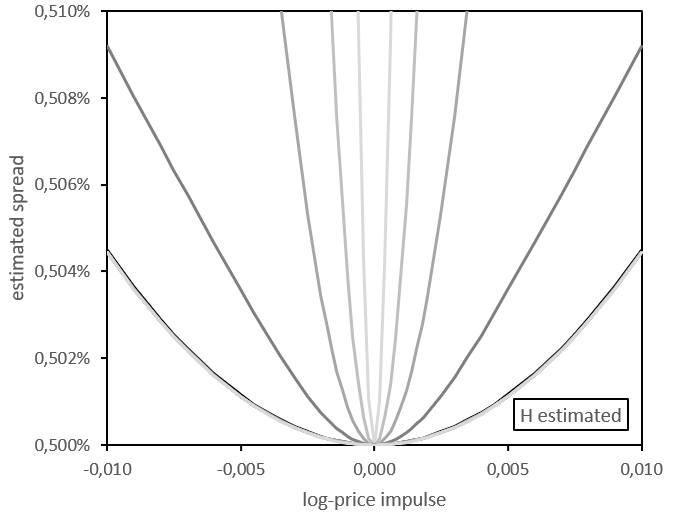} 
\begin{minipage}{0.7\textwidth}\caption{Estimated Hurst exponent (top left graph), estimated spread when $H$ is known (top right), when the estimated $H$ is plugged in the spread's estimator (bottom left and bottom right, detail), determined theoretically for an impulse in an observation. Each curve in the four graphs corresponds to a Hurst exponent between 0.1 to 0.9 (from bottom to top), with a step of 0.1 (the black curve is for $H=0.5$). Other parameters are $S=0.5\%$, $\sigma=0.2$, $\tau=1/(260\times 510)$, $L=1$, $n=60$ (observations in one hour).}
	\label{fig:Impulse}
\end{minipage}
\end{figure}

These different properties of the two estimators, $\widehat{S^2_{2,v}}(n,L,L',H)$ and $\widehat{S^2_{2,v}}(n,L,L',\widehat H_{L''})$, naturally raise the question of the sensitivity of the spread estimator to any divergence between the true Hurst exponent and its estimate. For this purpose, we present in Figure~\ref{fig:Impulse} the impulse response of the spread estimator. Starting from the theoretical variances of the price increments $V(\tau)$, $V(2\tau)$, and $V(4\tau)$, we consider an impulse $x$ for one of the observed log-prices. This leads to the response $2x^2/k_v(n,L)$ in the variance $\widehat V_v(n,L)$. Using equations~\eqref{eq:S2} and~\eqref{eq:estimH}, this leads to the responses displayed in Figure~\ref{fig:Impulse} for the estimators $\widehat H_{1}$, $\widehat{S^2_{2,v}}(n,1,2,H)$ and $\widehat{S^2_{2,v}}(n,1,2,\widehat H_{1})$. Interestingly, the response of the spread estimator is stronger for low values of $H$ when $H$ is known, but it is much stronger for high values of $H$ when $H$ is to be estimated. This will be confirmed in the simulation study, in Section~\ref{sec:simul}, in which one observes explosive estimates for the spread when both $H=0.7$ and $S>0.5\%$, whereas one does not observe such a phenomenon for $H=0.3$.

\subsubsection{Market model with autocorrelated trades}\label{sec:correlTrades}

When dealing with a serial dependence for the microstructure noise, the literature reports an exponentially decaying autocorrelation in high frequency~\cite[Section 2.1.3]{BBD}\cite{AMZ,JLZ}. To explain this autocorrelation, we can refer to a framework where a microstructure noise $\widetilde\eta_t$ is the discrete-valued counterpart of a hidden continuous process $\eta_t^{\star}$, using the sign function:
\begin{equation}\label{eq:OrnsteinBinarized}
\widetilde\eta_t=\sgn\left(\eta_t^{\star}\right)S.
\end{equation}
The exponential autocorrelation of $\widetilde\eta_t$ can stem from the specification of $\eta_t^{\star}$ as an Ornstein-Uhlenbeck process,
\begin{equation}\label{eq:OrnsteinSDE}
d\eta_t^{\star}=-\theta\eta_t^{\star} dt + \Xi dW_t,
\end{equation}
where $W_t$ is a standard Brownian motion, $\theta>0$ is the strength of the mean-reversion, and $\Xi>0$ is the magnitude of the diffusion. Proposition~\ref{pro:correlOU} provides the exact autocorrelation of such a binarized Ornstein-Uhlenbeck process.

\begin{pro}\label{pro:correlOU}
The autocorrelation of the process $\widetilde\eta_t$ introduced in equation~\eqref{eq:OrnsteinBinarized} is, for $0\leq s<t$:
$$\corr\left(\widetilde\eta_s,\widetilde\eta_t\right)=\frac{2}{\pi}\arctan\left(\left(e^{2\theta(t-s)}-1\right)^{-1/2}\right).$$
\end{pro}

The proof of Proposition~\ref{pro:correlOU} is postponed in Appendix~\ref{sec:correlOU}. A more practical expression is obtained  by applying a first-order Taylor expansion to the expression in Proposition~\ref{pro:correlOU}, which shows that the autocorrelation of the process $\widetilde\eta_t$ has asymptotically an exponential decay:
$$\corr\left(\widetilde\eta_s,\widetilde\eta_t\right)\overset{t-s\rightarrow \infty}{\sim}\frac{2}{\pi}e^{-\theta(t-s)}.$$
This legitimizes working with a microstructure noise $\eta_t$ having an exponential autocorrelation, as exposed in Assumption~\ref{assum:correlTrades}.

\begin{assum}\label{assum:correlTrades}
We assume that the only serial dependence is for the microstructure noise, with an exponentially decaying autocorrelation:
\begin{itemize}
\item $p_t^{\star}$ is a Brownian motion with volatility parameter $\sigma>0$,
\item $\eta_s$ and $\eta_t$ are correlated, with a correlation $\exp\left(-\frac{|t-s|}{\lambda}\right)$, where $\lambda>0$.
\end{itemize}
\end{assum}

Using Assumptions~\ref{assum:indepProcesses} and~\ref{assum:correlTrades}, the theoretical variance of the price log-returns becomes, whatever $\tau\geq 0$,
\begin{equation}\label{eq:varianceIncrPrix_correlTrades}
V(L\tau)=\Var(p_{t+L\tau}-p_t)=L\tau\sigma^2 + \frac{S^2}{2}\left(1-\exp\left(-\frac{L\tau}{\lambda}\right)\right),
\end{equation}
where $L\in\llbracket 1,n\rrbracket$. Like in Section~\ref{sec:fBm}, combining several moments can make $\sigma^2$ disappear, leading to an estimator of $S^2$, for $v\in\{1,2,3\}$, which depends on the other parameter of the model, $\lambda$, or on the convenient notation $\rho=e^{-\tau/\lambda}$:
\begin{equation}\label{eq:S3}
\widehat{S^2_{3,v}}(n,L,L',\rho)=\frac{2\left(L'\widehat V_v(n,L)-L\widehat V_v(n,L')\right)}{L'\left(1-\rho^L\right)-L\left(1-\rho^{L'}\right)}.
\end{equation}
Like in Section~\ref{sec:fBm}, we can show that forcing $\rho=0$ in equation~\eqref{eq:S3} leads to the standard estimator $\widehat{S^2_{1,v}}(n,L,L')$ introduced in equation~\eqref{eq:S1}. But, if trades are really correlated, introducing the estimator of equation~\eqref{eq:S3} is required since the standard estimator $\widehat{S^2_{1,v}}(n,L,L')$ is biased. For $L=1$ and $L'=2$, this bias is equal to $-S^2(2\rho-\rho^2)\leq 0$ and is the same for Roll's estimator. We will see in the simulation study that $\widehat{S^2_{3,v}}(n,L,L',\rho)$ is the only tested estimator that does not underestimate $S^2$.

In practice however, $\lambda$ and thus $\rho$ are unknown. Considering that
$$\left\{\begin{array}{cclcl}
2V(L\tau)-V(2L\tau) & = & \frac{S^2}{2}\left(1-\rho^L\right)^2 & & \\
2V(2L\tau)-V(4L\tau) & = & \frac{S^2}{2}\left(1-\rho^{2L}\right)^2 & = & \frac{S^2}{2}\left(1-\rho^L\right)^2\left(1+\rho^L\right)^2, 
\end{array}\right.$$
we obtain the following estimator of $\rho^L$:
\begin{equation}\label{eq:estimcorrel}
\widehat{\rho^L}=\sqrt{\left|\frac{2\widehat V_v(n,2L)-\widehat V_v(n,4L)}{2\widehat V_v(n,L)-\widehat V_v(n,2L)}\right|}-1,
\end{equation}
where the absolute value is only intended to keep the square root defined. Plugging equation~\eqref{eq:estimcorrel} into equation~\eqref{eq:S3} leads to an estimator $\widehat{S^2_{3,v}}(n,L,L',(\widehat{\rho^{L''}})^{1/L''})$ of $S^2$ that does not depend on unknown parameters of the model. The most natural choice for $L'$ and $L''$ is $(L',L'')=(2L,L)$ leading to
\begin{equation}\label{eq:S3_correlUnknown}
\widehat{S^2_{3,v}}(n,L,2L,(\widehat{\rho^L})^{1/L})=\frac{2\left(2\widehat V_v(n,L)-\widehat V_v(n,2L)\right)^2}{\left(2\sqrt{\left|2\widehat V_v(n,L)-\widehat V_v(n,2L)\right|}-\sqrt{\left|2\widehat V_v(n,2L)-\widehat V_v(n,4L)\right|}\right)^2}.
\end{equation}

In order to determine the asymptotic properties of $\widehat{S^2_{3,v}}(n,L,L',\rho)$ and $\widehat{S^2_{3,v}}(n,L,L',(\widehat{\rho^{L''}})^{1/L''})$, we study the function $K$, introduced in equation~\eqref{eq:covIncrSq}, and the first two moments of the empirical variance estimators. This is the purpose of Propositions~\ref{pro:covIncrSq3} and~\ref{pro:momentV3}.

\begin{pro}\label{pro:covIncrSq3}
Under the market model introduced in equation~\eqref{eq:logprice} and Assumptions~\ref{assum:indepProcesses} and~\ref{assum:correlTrades}, the covariance $K(u,\delta)$ introduced in equation~\eqref{eq:covIncrSq}, for $u\geq 0$, is such that
$$K(u,\delta)=\left\{\begin{array}{ll}
V(u)^2 & \text{if } \delta\leq 0 \\
V(u)^2 + 2\delta^2\sigma^4 + \frac{S^4}{4}\left(e^{-2(u-\delta)/\lambda}-e^{-2u/\lambda}\right) & \\
+ \delta \sigma^2S^2\left(2e^{-(u-\delta)/\lambda}-e^{-\delta/\lambda}-e^{-(2u-\delta)/\lambda}\right) & \text{if } \delta\in[0,u],
\end{array}\right.$$
where $V$ is defined in equation~\eqref{eq:varianceIncrPrix_correlTrades}.
\end{pro}

The proof of Proposition~\ref{pro:covIncrSq3} is postponed in Appendix~\ref{sec:covIncrSq3}.

We note that despite the non-zero autocorrelation between the trades, $K(u,\delta)$ does not depend on $\delta$ when $\delta\leq 0$. It is a consequence of the short-range dependence of the process, stemming from the exponential decay of the autocorrelation of the trades.

\begin{pro}\label{pro:momentV3}
Under the market model introduced in equation~\eqref{eq:logprice} and Assumptions~\ref{assum:indepProcesses} and~\ref{assum:correlTrades}, we have
$$\E\left[\widehat V_v(n,L)\right]=V(L\tau)$$
for all $v\in\{1,2,3\}$, where $V$ is defined in equation~\eqref{eq:varianceIncrPrix_correlTrades}. Moreover, the variance of the empirical variance estimator is $\Var\left[\widehat V_1(n,L)\right]=\sigma^2_{3,1}(L)/k_1(n,L) + \mathcal{O}\left(n^{-2}\right)$, where
$$\begin{array}{ccl}
\sigma^2_{3,1}(L) & = & 2L^2\tau^2\sigma^4 + \frac{S^4}{4}\left(1-e^{-2L\tau/\lambda}\right) + 2L\tau \sigma^2S^2\left(1-e^{-L\tau/\lambda}\right) \\
 & & + \frac{2}{3}\tau^2\sigma^4(L-1)L(2L-1) + \frac{S^4}{2}e^{-2L\tau/\lambda}\left(f_L\left(\frac{2\tau}{\lambda}\right)-L+1\right)  \\
 & & + 2\tau \sigma^2S^2\left(2e^{-L\tau/\lambda}f_L'\left(\frac{\tau}{\lambda}\right) - f_L'\left(-\frac{\tau}{\lambda}\right) -e^{-2L\tau/\lambda}f_L'\left(\frac{\tau}{\lambda}\right)\right),
\end{array}$$
and where $f_L(x)=(e^x-e^{xL})/(1-e^x)$ and $f_L'(x)=(e^{x}-Le^{xL}-e^{x(L+1)})/(1-e^x)^2$. If $v\in\{2,3\}$, we have $\Var\left[\widehat V_v(n,L)\right]=\sigma^2_{3,v}(L)/k_v(n,L)$, where
$$\sigma^2_{3,v}(L)=2L^2\tau^2\sigma^4 + \frac{S^4}{4}\left(1-e^{-2L\tau/\lambda}\right)
+ 2L\tau \sigma^2S^2\left(1-e^{-L\tau/\lambda}\right).$$
\end{pro}

The proof of Proposition~\ref{pro:momentV3} is postponed in Appendix~\ref{sec:momentV3}. When the autocorrelation of the trades is negligible, that is when $\lambda\rightarrow 0$, we obtain Proposition~\ref{pro:momentV1} as a limit case of Proposition~\ref{pro:momentV3}. Finally, Theorem~\ref{thm:S3} gathers the asymptotic properties of the estimator of $S^2$, which depend on whether the parameter $\lambda$ is known or estimated.

\begin{thm}\label{thm:S3}
Under the market model introduced in equation~\eqref{eq:logprice} and Assumptions~\ref{assum:indepProcesses} and~\ref{assum:correlTrades}, the statistics $\widehat{S^2_{3,v}}(n,L,L',\rho)$ and $\widehat{S^2_{3,v}}(n,L,L',(\widehat{\rho^{L''}})^{1/L''})$ introduced in equations~\eqref{eq:S3} and~\eqref{eq:estimcorrel}, for $v\in\{1,2,3\}$ and $L$, $L'$, and $L''$ fixed integers such that $L'\neq L$, are consistent consistent estimators of $S^2$. In addition, $\widehat{S^2_{3,v}}(n,L,L',\rho)$ is unbiased. Moreover, 
\begin{equation}\label{eq:TCL3}
\sqrt{n}\left(\widehat{S^2_{3,v}}(n,L,L',\rho) - S^2\right)\overset{d}{\longrightarrow} \mathcal N\left(0,\gamma_{3,v}(L,L',\rho)\right),
\end{equation}
where
$$\begin{array}{ccl}
\gamma_{3,v}(L,L',\rho) & = & \frac{4}{(L'(1-\rho^L)-L(1-\rho^{L'}))^2} \Bigl[L'^2\zeta_v(L)\sigma_{3,v}^2(L) + L^2\zeta_v(L')\sigma_{3,v}^2(L') \\
 & & - 2LL'\sqrt{\zeta_v(L)\zeta_v(L')}\sigma_{3,v}(L)\sigma_{3,v}(L')r_{3,v}(L,L')\Bigr],
\end{array}$$
with $\sigma_{3,v}(L)$ defined as in Proposition~\ref{pro:momentV3}, $r_{3,v}(L,L')=\lim_{n\rightarrow\infty}\corr\left(\widehat V_v(n,L),\widehat V_v(n,L')\right)$, and $(\zeta_1(L),\zeta_2(L),\zeta_3(L))$ defined as in Theorem~\ref{thm:S1}. We also have, for $\rho>0$,
\begin{equation}\label{eq:TCL3_correlUnknown}
\sqrt{n}\left(\widehat{S^2_{3,v}}(n,L,2L,(\widehat{\rho^L})^{1/L}) - S^2\right)\overset{d}{\longrightarrow} \mathcal N\left(0,\mathbf W^T\Sigma\mathbf W\right),
\end{equation}
where, for $i,j\in\llbracket 1,3\rrbracket$, $\Sigma_{ij}=\sqrt{\zeta_v(2^{i-1}L)\zeta_v(2^{j-1}L)}\sigma_{3,v}(2^{i-1}L)\sigma_{3,v}(2^{j-1}L)r_{3,v}(2^{i-1}L,2^{j-1}L)$, $\mathbf W = \nabla h(V(L\tau),V(2L\tau),V(4L\tau))$, and 
\begin{equation}\label{eq:deltafunction3}
h:(x,y,z)\mapsto \frac{2(2x-y)^2}{(2\sqrt{2x-y}-\sqrt{2y-z})^2}.
\end{equation}
\end{thm}

The proof of Theorem~\ref{thm:S3} is postponed in Appendix~\ref{sec:S3}. Since the microstructure noise is $\alpha$-mixing, it is very close to the proof of Theorem~\ref{thm:S1}, with a multivariate central limit theorem and an application of the delta method. Like in Theorem~\ref{thm:S2}, for the model with autocorrelated price increments, our estimator of $S^2$ is only unbiased when the dependence parameter, that is $\rho$ or $\lambda$, is known. In practice, we estimate it, and the estimator of $S^2$ is then only asymptotically unbiased.

\subsubsection{Fully autocorrelated market model}\label{sec:fBmAndCorrelTrades}

Combining the two kinds of serial dependence introduced in Sections~\ref{sec:fBm} and~\ref{sec:correlTrades}, we obtain a more general and realistic framework.

\begin{assum}\label{assum:fBmAndCorrelTrades}
We assume that there is a serial dependence both for the increments of $p_t^{\star}$ and for $\eta_t$:
\begin{itemize}
\item $p_t^{\star}=\sigma B^H_t$, where $\sigma>0$ and $B^H_t$ is an fBm of Hurst exponent $H\in(0,1)$,
\item $\eta_s$ and $\eta_t$ are correlated, with a correlation $\exp\left(-\frac{|t-s|}{\lambda}\right)$, where $\lambda>0$.
\end{itemize}
\end{assum}

Using Assumptions~\ref{assum:indepProcesses} and~\ref{assum:fBmAndCorrelTrades}, we obtain the theoretical variance of the price log-returns,
$$V(L\tau)=\Var(p_{t+L\tau}-p_t)=(L\tau)^{2H}\sigma^2 + \frac{S^2}{2}\left(1-\exp\left(-\frac{L\tau}{\lambda}\right)\right),$$
for $\tau\geq 0$ and $L\in\llbracket 1,n\rrbracket$. The parameters $S$, $\sigma^2$, $H$, and $\lambda$ all appear in the equation and are not directly observed. If one uses the standard estimator introduced in equation~\eqref{eq:S1}, there will obviously be a bias and a specific estimator is thus required.

Like in the three other frameworks introduced above, it may be possible to get rid of $\sigma^2$ by combining variances of several time scales. One might also define estimators for $H$ and $\lambda$, leading to an estimator of $S$ based on these intermediate estimators. To avoid the problems induced by this errors-in-variables approach, we instead propose to estimate the four parameters together with a numerical optimization, consisting in the minimization of $(V(l\tau)-\widehat V_v(n,l))^2$ for various resolutions $l$:
$$\left(\widehat{S^2_{4,v}},\widehat{\sigma^2_{4,v}},\widehat{H_{4,v}},\widehat{\lambda_{4,v}}\right)(n,L) = \underset{S^2\geq 0,\sigma^2> 0,H\in(0,1),\lambda> 0}{\argmin} \sum_{l=1}^L\left[(L\tau)^{2H}\sigma^2 + \frac{S^2}{2}\left(1-e^{-L\tau/\lambda}\right)-\widehat V_v(n,l)\right]^2.$$

\section{An application to data}\label{sec:simul}

We now study the estimators introduced in Section~\ref{sec:estimators} in various simulated market situations: without serial dependence, with correlated increments of the mid price, or with correlated trades. These estimators will be compared to other existing estimators, which have not been designed to tackle serial dependence. The case of infrequent trading will also be considered. Last, we also consider a short application to real data.

\subsection{Alternative estimation methods}\label{sec:altEstim}

In his pioneering work on bid-ask spread estimation, Roll proposes an estimator of $S^2$ based on the covariance of successive price increments~\cite{Roll}:
$$\text{Roll}^2=-\frac{4}{n-2}\sum_{i=0}^{n-3}\left(p_{(i+1)\tau}-p_{i\tau}\right)\left(p_{(i+2)\tau}-p_{(i+1)\tau}\right),$$
where $p_t$ is the close price in an interval $[t-\tau,t]$, like in Section~\ref{sec:estim}. This estimator of the squared spread may be negative and one can use a classical transformation, such as the one we use in this section for the estimators introduced in Section~\ref{sec:estimators}: $\widehat S=\max(0,\text{Roll}^2)^{1/2}$. Roll's estimator is based on an independence of the mid price and the trades, that is Assumption~\ref{assum:indepProcesses}, as well as on an absence of serial correlation regarding $p_t^{\star}$. Indeed, with these assumptions, $\Cov(p_{t+\tau}-p_{t},p_{t+2\tau}-p_{t+\tau})=-S^2/4$. Roll's hypothesis are more general than Assumption~\ref{assum:zeroCorrel}, since they don't require that successive increments are independent, but only uncorrelated. In fact, our estimators are still consistent if the independence is replaced by an absence of correlation in Assumption~\ref{assum:zeroCorrel}, but our stricter framework is useful for obtaining an asymptotic distribution of estimators.

Adding the Gaussian hypothesis, that is working exactly with Assumptions~\ref{assum:indepProcesses} and~\ref{assum:zeroCorrel}, Corwin and Schultz propose an approach not based on moments but on the distribution of the range covered by a Brownian motion in a given time interval~\cite{CS}. The difference between this estimator and Roll's is of the same kind, in the most common framework of volatility estimation, as the difference between Parkinson volatility~\cite{Parkinson,Molnar} and realised bipower variation~\cite{BNS,BNGJ}. For the time interval $[t,t+\tau]$, we define the high and low prices: $h_{t,t+\tau}=\max_{s\in[t,t+\tau]}p_s$ and $l_{t,t+\tau}=\min_{s\in[t,t+\tau]}p_s$. We also introduce $o_t$, the corresponding open price, which will be useful later for AGK$_1$ estimator. One then considers the squared high-low difference for two time intervals:
$$\left\{\begin{array}{ccl}
\beta_{t} & = & \left(h_{t,t+\tau}-l_{t,t+\tau}\right)^2 + \left(h_{t+\tau,t+2\tau}-l_{t+\tau,t+2\tau}\right)^2 \\
\gamma_{t} & = & \left(h_{t,t+2\tau}-l_{t,t+2\tau}\right)^2. \\
\end{array}\right.$$
Then, thanks to a numerical optimization, one solves the following equation in $\epsilon_t$:
$$0=\epsilon_{t}^2\left(2\kappa_2^2(1-\sqrt{2})+\kappa_1\right)+ \epsilon_t 2\kappa_2(\sqrt{2}-1)\sqrt{\epsilon_t^2(\kappa_2^2-\kappa_1)+\frac{\beta_t}{2}} +\frac{\beta_t}{2}-\gamma_t,$$
where $\kappa_1=4\ln(2)$ and $\kappa_2=\sqrt{8/\pi}$. Then, noting $\alpha_t= -\kappa_2 \epsilon_{t}+\sqrt{\epsilon_{t}^2(\kappa_2^2-\kappa_1)+\frac{\beta_{t}}{2}}$, one obtains the following estimator for the spread, in the interval $[t,t+2\tau]$:
$$\text{CS}_t=\frac{2\left(e^{\alpha_t}-1\right)}{1+e^{\alpha_t}}.$$
One can then average these estimators to cover all the data available in a day:
$$\text{CS}=\frac{1}{n-2}\sum_{i=0}^{n-3}\text{CS}_{i\tau}.$$

However, overnight effects may bias this estimator. To filter these effects, Abdi and Ranaldo propose an estimator extending the two previous ones~\cite{AR}. Like CS, it is based on the high-low range, more precisely on the mid-range $m_{t,t+\tau}=(h_{t,t+\tau}+l_{t,t+\tau})/2$. Like Roll's estimator, the spread is obtained as a covariance of price increments in successive intervals $[t,t+\tau]$ and $[t+\tau,t+2\tau]$, where one does not consider the close but either the mid-range price or the close:
$$\text{AR}^2=-\frac{4}{n-2}\sum_{i=0}^{n-3}\left(p_{(i+1)\tau}-m_{i\tau,(i+1)\tau}\right)\left(m_{(i+1)\tau,(i+2)\tau}-p_{(i+1)\tau}\right).$$
This estimator of $S^2$ relies on Assumptions~\ref{assum:indepProcesses} and~\ref{assum:zeroCorrel} and thus includes the Gaussian aspect, unlike Roll's.

In a more recent work, an estimator based on a covariance, like Roll and AR, has been proposed to handle infrequent trading~\cite{AGK}. Indeed, if there is no trade between two consecutive observation dates, the corresponding prices are on the same side (bid or ask) and this creates a spurious correlation of the microstructure noise. The proposed estimator is then
$$\text{AGK}_1^2=\frac{-8\sum_{i=0}^{n-3}\left(\overline{m_{(i+1)\tau,(i+2)\tau}-o_{(i+1)\tau}}\right)\left(o_{(i+1)\tau}-p_{i\tau}\right)}{\sum_{i=0}^{n-3}\left(\indic_{o_{(i+1)\tau}\neq h_{(i+1)\tau,(i+2)\tau}} \indic_{o_{(i+1)\tau}\neq l_{(i+1)\tau,(i+2)\tau}}\right)\left(1-\indic_{p_{i\tau}=h_{(i+1)\tau,(i+2)\tau}=l_{(i+1)\tau,(i+2)\tau}}\right)},$$  
where the denominator is a correction term in case of infrequent or non-existent trading and where $\overline{x}$ is a centred version of the variable $x$, with the mean calculated only on time intervals with frequent trading. In the simulations of Section~\ref{sec:resSimul}, we have considered continuous trading, but the framework of infrequent trading is also studied in Section~\ref{sec:infreq}.

In the same article, the authors propose an other estimator for $S$, combining in a more complex way, open, high, low, and close prices~\cite{AGK}. The authors call this estimator EDGE (efficient discrete generalized estimator) and we will note it $\text{AGK}_2$ in what follows.

\subsection{Comparison of the methods on simulated data}\label{sec:resSimul}

We simulate 10,000 daily price trajectories of a stock. Each day lasts 8 hours and consists of 28,800 seconds. The price, simulated each second, follows equation~\eqref{eq:logprice} and is thus the sum of a Brownian motion (model 1) or an fBm (model 2) with a microstructure noise. The volatility parameter at a daily time scale is $3\%$. The microstructure noise is a series of independent variables, except in model 3, where it derives from an Ornstein-Uhlenbeck process, like in equations~\eqref{eq:OrnsteinBinarized} and~\eqref{eq:OrnsteinSDE}. We note that this last model leads to a non-asymptotic correlation structure which is slightly different from Assumption~\ref{assum:correlTrades}, but, as will be showed, this does not prevent the corresponding estimator to perform well in this situation. The parameter $\theta$ in this model is set to 0.01 at a one-second time scale. This means that the correlation between two trades falls below $50\%$ for a lag of 35 seconds.

From this time series of prices generated at a one-second frequency, we build time series sampled every minute. This means that, for each non-overlapping one-minute interval, corresponding to the duration $\tau$, we determine the open, high, low, and close prices from the simulated high-frequency time series, observed with time steps $\tau/60$. The estimators put forward in Section~\ref{sec:estimators} only use the 480 one-minute close prices generated for each trajectory. 

In what follows, we observe the output of four estimators introduced in Section~\ref{sec:estimators}, $\widehat S_{1,1}$, $\widehat S_{2,1}$, $\widehat S_{3,1}$, and $\widehat S_{4,1}$, in the version where the parameters $H$ and $\lambda$ are estimated for $\widehat S_{2,1}$ and $\widehat S_{3,1}$, as well as the four benchmark estimators presented in Section~\ref{sec:altEstim}. Thanks to the 10,000 simulated trajectories, we can display the bias, standard deviation, and quadratic risk related to each estimator. 

We also make a bootstrap-based statistical test, to determine whether the true spread value $S$ is plausible according to the empirical distribution of estimated spreads. More precisely, we use a Gaussian distribution for the estimated spread, as asymptotically suggested by Section~\ref{sec:estim}. The mean and variance of this distribution, for each estimator, are respectively the average and the empirical variance of the estimates. The null hypothesis is that, under this distribution based on estimates, the true spread is indeed $S$. If it is rejected, it means that the true spread is an outlier in the distribution of the estimates, so that the estimator is unable to find the true spread or leads to an erroneous perception of what the true spread is. Relevant estimators have to accept the null hypothesis.

We first study model 1, the baseline situation in which there is no serial dependence. The results are gathered in Table~\ref{tab:simul_1}. Among our estimators, the most (respectively least) accurate, regarding both the bias and the variance, is $\widehat{S}_{1,1}$ (resp. $\widehat{S}_{4,1}$). This result is not surprising since $\widehat{S}_{1,1}$ has specifically been design for this model. Among the benchmark estimators, Roll has the lowest bias, AGK$_2$ the lowest variance, and AR the lowest quadratic risk. All the estimators are relevant, according to our statistical test, except AGK$_2$ and even more CS. These estimators lead to an overconfidence in the estimates because of a significant bias, with respect to the small variance. The same property is observed for AGK$_2$ and CS whatever the model evaluated in our study.

\begin{table}[htbp]
\centering
\begin{tabular}{|l|c|c|c|c|}
\hline
Estimator & Bias & Standard dev. & Quadratic risk & test spread = $S$ (p-value) \\
\hline
CS & $-3.2\times 10^{-4}$ & $6.0\times 10^{-5}$ & $1.0\times 10^{-7}$ & Rejected ($1.1\times 10^{-7}$) \\ 
AR & $-9.2\times 10^{-6}$ & $5.1\times 10^{-5}$ & $\mathbf{2.6\times 10^{-9}}$ & \textbf{Accepted} ($8.6\times 10^{-1}$) \\
Roll & $\mathbf{-6.6\times 10^{-7}}$ & $2.9\times 10^{-4}$ & $8.2\times 10^{-8}$ & \textbf{Accepted} ($1.0\times 10^{0}$) \\ 
AGK$_1$ & $1.2\times 10^{-4}$ & $1.3\times 10^{-4}$ & $3.1\times 10^{-8}$ & \textbf{Accepted} ($3.8\times 10^{-1}$) \\
AGK$_2$ & $1.2\times 10^{-4}$ & $\mathbf{4.3\times 10^{-5}}$ & $1.6\times 10^{-8}$ & Rejected ($4.7\times 10^{-3}$) \\  
\hline
$\widehat S_{1,1}$ & $\mathbf{3.9\times 10^{-6}}$ & $\mathbf{1.8\times 10^{-4}}$ & $\mathbf{3.1\times 10^{-8}}$ & \textbf{Accepted} ($9.8\times 10^{-1}$) \\
$\widehat S_{2,1}$ & $-1.8\times 10^{-4}$ & $5.7\times 10^{-4}$ & $3.5\times 10^{-7}$ & \textbf{Accepted} ($7.5\times 10^{-1}$) \\
$\widehat S_{3,1}$ & $3.1\times 10^{-4}$ & $9.2\times 10^{-4}$ & $9.5\times 10^{-7}$ & \textbf{Accepted} ($7.3\times 10^{-1}$) \\
$\widehat S_{4,1}$ & $-1.0\times 10^{-3}$ & $1.9\times 10^{-3}$ & $4.8\times 10^{-6}$ & \textbf{Accepted} ($5.9\times 10^{-1}$) \\
\hline
\end{tabular}
\begin{minipage}{0.9\textwidth}\caption{For $S=0.5\%$ and 10,000 simulated trajectories of one trading day using model 1, bias, standard deviation, and quadratic risk of the estimators. The last column is the output of the bootstrap statistical test of relevance of the spread, with a confidence of $95\%$ (corresponding p-value in parenthesis).}
\label{tab:simul_1}
\end{minipage}
\end{table}

We now study the reaction of estimators to the presence of serial dependence in the increments of mid prices. We consider negative and positive autocorrelations, namely a Hurst exponent of 0.3 in Table~\ref{tab:simul_2H3} and 0.7 in Table~\ref{tab:simul_2H7}. As explained in Section~\ref{sec:fBm}, we can see in the tables that most of the estimators tend to overestimate (respectively underestimate) the true spread when $H<1/2$ (resp. $H>1/2$). When $H=0.3$, the estimator we have introduced to face this situation, $\widehat S_{2,1}$, is an exception and it even has the smallest bias. As discussed in the study of the impulse response in Section~\ref{sec:fBm}, it is less performing for $H=0.7$. Overall, AGK$_1$ (for $H=0.3$), AR (for $H=0.7$), and, among our estimators and whatever $H\in\{0.3,0.7\}$, $\widehat S_{1,1}$ have the lowest quadratic risk. Nevertheless, most estimators have to be discarded for $H=0.3$, according to the statistical test, which shows that only $\widehat S_{2,1}$, $\widehat S_{4,1}$, and AGK$_1$ are relevant.

\begin{table}[htbp]
\centering
\begin{tabular}{|l|c|c|c|c|}
\hline
Estimator & Bias & Standard dev. & Quadratic risk & test spread = $S$ (p-value) \\
\hline
CS & $2.0\times 10^{-3}$ & $1.7\times 10^{-4}$ & $4.1\times 10^{-6}$ & Rejected ($<10^{-10}$) \\ 
AR & $1.0\times 10^{-3}$ & $2.2\times 10^{-4}$ & $1.1\times 10^{-6}$ & Rejected ($5.7\times 10^{-6}$) \\
Roll & $1.8\times 10^{-3}$ & $5.2\times 10^{-4}$ & $3.5\times 10^{-6}$ & Rejected ($4.7\times 10^{-4}$) \\
AGK$_1$ & $\mathbf{3.4\times 10^{-4}}$ & $2.4\times 10^{-4}$ & $\mathbf{1.8\times 10^{-7}}$ & \textbf{Accepted} ($1.5\times 10^{-1}$) \\
AGK$_2$ & $6.8\times 10^{-4}$ & $\mathbf{1.4\times 10^{-4}}$ & $4.8\times 10^{-7}$ & Rejected ($1.6\times 10^{-6}$) \\  
\hline
$\widehat S_{1,1}$ & $2.2\times 10^{-3}$ & $\mathbf{3.7\times 10^{-4}}$ & $\mathbf{5.1\times 10^{-6}}$ & Rejected ($1.8\times 10^{-9}$) \\
$\widehat S_{2,1}$ & $\mathbf{-1.3\times 10^{-4}}$ & $2.4\times 10^{-3}$ & $5.9\times 10^{-6}$ & \textbf{Accepted} ($9.6\times 10^{-1}$) \\
$\widehat S_{3,1}$ & $6.3\times 10^{-3}$ & $2.2\times 10^{-3}$ & $4.4\times 10^{-5}$ & Rejected ($4.7\times 10^{-3}$) \\
$\widehat S_{4,1}$ & $9.6\times 10^{-4}$ & $2.5\times 10^{-3}$ & $7.3\times 10^{-6}$ & \textbf{Accepted} ($7.0\times 10^{-1}$) \\
\hline
\end{tabular}
\begin{minipage}{0.9\textwidth}\caption{For $S=0.5\%$, $H=0.3$, and 10,000 simulated trajectories of one trading day using model 2, bias, standard deviation, and quadratic risk of the estimators. The last column is the output of the bootstrap statistical test of relevance of the spread, with a confidence of $95\%$ (corresponding p-value in parenthesis).}
\label{tab:simul_2H3}
\end{minipage}
\end{table}

\begin{table}[htbp]
\centering
\begin{tabular}{|l|c|c|c|c|}
\hline
Estimator & Bias & Standard dev. & Quadratic risk & test spread = $S$ (p-value) \\
\hline
CS & $-2.7\times 10^{-4}$ & $2.5\times 10^{-5}$ & $7.2\times 10^{-8}$ & Rejected ($<10^{-10}$) \\ 
AR & $\mathbf{-8.9\times 10^{-6}}$ & $\mathbf{1.1\times 10^{-5}}$ & $\mathbf{2.0\times 10^{-10}}$ & \textbf{Accepted} ($4.1\times 10^{-1}$) \\
Roll & $-2.3\times 10^{-5}$ & $2.6\times 10^{-4}$ & $6.8\times 10^{-8}$ & \textbf{Accepted} ($9.3\times 10^{-1}$) \\
AGK$_1$ & $2.1\times 10^{-4}$ & $1.2\times 10^{-4}$ & $5.9\times 10^{-8}$ & \textbf{Accepted} ($9.2\times 10^{-2}$) \\
AGK$_2$ & $2.1\times 10^{-4}$ & $2.1\times 10^{-5}$ & $4.3\times 10^{-8}$ & Rejected ($<10^{-10}$) \\  
\hline
$\widehat S_{1,1}$ & $\mathbf{-4.3\times 10^{-5}}$ & $\mathbf{1.4\times 10^{-4}}$ & $\mathbf{2.2\times 10^{-8}}$ & \textbf{Accepted} ($7.6\times 10^{-1}$) \\
$\widehat S_{2,1}$ & $-2.8\times 10^{-4}$ & $1.0\times 10^{-3}$ & $1.1\times 10^{-6}$ & \textbf{Accepted} ($7.8\times 10^{-1}$) \\
$\widehat S_{3,1}$ & $-3.5\times 10^{-4}$ & $3.5\times 10^{-4}$ & $2.4\times 10^{-7}$ & \textbf{Accepted} ($3.1\times 10^{-1}$) \\
$\widehat S_{4,1}$ & $-9.1\times 10^{-4}$ & $1.6\times 10^{-3}$ & $3.3\times 10^{-6}$ & \textbf{Accepted} ($5.7\times 10^{-1}$) \\
\hline
\end{tabular}
\begin{minipage}{0.9\textwidth}\caption{For $S=0.5\%$, $H=0.7$, and 10,000 simulated trajectories of one trading day using model 2, bias, standard deviation, and quadratic risk of the estimators. The last column is the output of the bootstrap statistical test of relevance of the spread, with a confidence of $95\%$ (corresponding p-value in parenthesis).}
\label{tab:simul_2H7}
\end{minipage}
\end{table}

In model 3, the trades are positively correlated. As one can see in Table~\ref{tab:simul_3}, this leads to a negative bias. This underestimation is thus observed for any kind of positive serial correlation, that is for both trades and increments of mid prices. In Table~\ref{tab:simul_3}, $\widehat S_{3,1}$ is the only exception, with a slightly positive bias, which is also the lowest in absolute value. It also has the lowest quadratic risk among all the estimators. Regarding the statistical test, all the estimators are rejected, except $\widehat S_{2,1}$, $\widehat S_{3,1}$, and $\widehat S_{4,1}$. Among the benchmark estimators, the covariance-based Roll estimator is the least inaccurate.

\begin{table}[htbp]
\centering
\begin{tabular}{|l|c|c|c|c|}
\hline
Estimator & Bias & Standard dev. & Quadratic risk & test spread = $S$ (p-value) \\
\hline
CS & $-3.1\times 10^{-3}$ & $\mathbf{1.3\times 10^{-4}}$ & $9.8\times 10^{-6}$ & Rejected ($<10^{-10}$) \\ 
AR & $-2.0\times 10^{-3}$ & $1.4\times 10^{-4}$ & $3.9\times 10^{-6}$ & Rejected ($<10^{-10}$) \\
Roll & $\mathbf{-1.6\times 10^{-3}}$ & $3.1\times 10^{-4}$ & $\mathbf{2.8\times 10^{-6}}$ & Rejected ($2.0\times 10^{-7}$) \\
AGK$_1$ & $-3.5\times 10^{-3}$ & $1.8\times 10^{-4}$ & $1.3\times 10^{-5}$ & Rejected ($< 10^{-10}$) \\
AGK$_2$ & $-2.9\times 10^{-3}$ & $2.0\times 10^{-4}$ & $8.2\times 10^{-6}$ & Rejected ($< 10^{-10}$) \\  
\hline
$\widehat S_{1,1}$ & $-1.3\times 10^{-3}$ & $\mathbf{2.2\times 10^{-4}}$ & $1.6\times 10^{-6}$ & Rejected ($2.3\times 10^{-8}$) \\
$\widehat S_{2,1}$ & $-2.4\times 10^{-3}$ & $1.3\times 10^{-3}$ & $7.4\times 10^{-6}$ & \textbf{Accepted} ($7.9\times 10^{-2}$) \\
$\widehat S_{3,1}$ & $\mathbf{3.3\times 10^{-4}}$ & $1.1\times 10^{-3}$ & $\mathbf{1.4\times 10^{-6}}$ & \textbf{Accepted} ($7.7\times 10^{-1}$) \\
$\widehat S_{4,1}$ & $-1.7\times 10^{-3}$ & $2.0\times 10^{-3}$ & $6.9\times 10^{-6}$ & \textbf{Accepted} ($3.7\times 10^{-1}$) \\
\hline
\end{tabular}
\begin{minipage}{0.9\textwidth}\caption{For $S=0.5\%$ and 10,000 simulated trajectories of one trading day using model 3, bias, standard deviation, and quadratic risk of the estimators. The last column is the output of the bootstrap statistical test of relevance of the spread, with a confidence of $95\%$ (corresponding p-value in parenthesis).}
\label{tab:simul_3}
\end{minipage}
\end{table}

In practical applications, determining whether there is serial dependence or not and whether this serial dependence applies to the series of mid-price increments or to the microstructure noise requires an additional statistical work. But our estimator obtained by numerical optimization, $\widehat S_{4,1}$, seems to behave satisfyingly, whatever the model, according to all the above tables, since the statistical test does not lead to its rejection.

In all the above analysis, we have considered a true spread equal to $0.5\%$. We now study the sensitivity of the results to a variation of this true spread. For models 1, 2, and 3, we analyse the simulated bias and standard deviation of several estimators for spreads equal to $0.1\%$, $0.25\%$, $0.5\%$, $0.75\%$, and $1\%$.

\begin{figure}[htbp]
	\centering
		\includegraphics[width=0.45\textwidth]{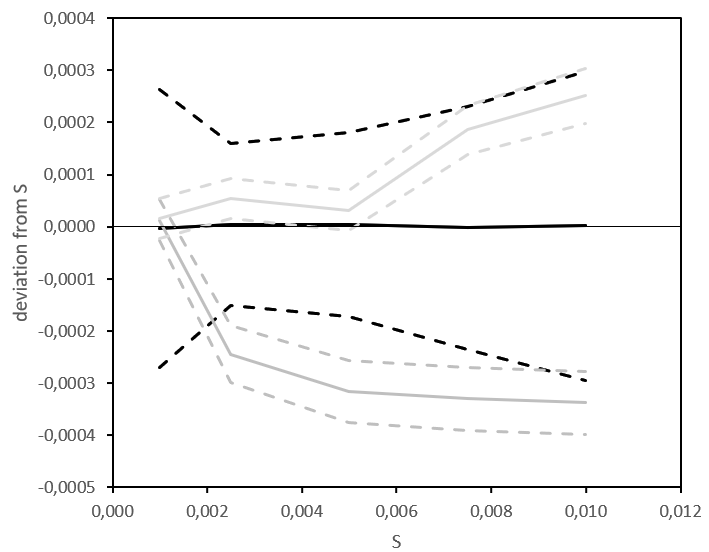} 
		\includegraphics[width=0.45\textwidth]{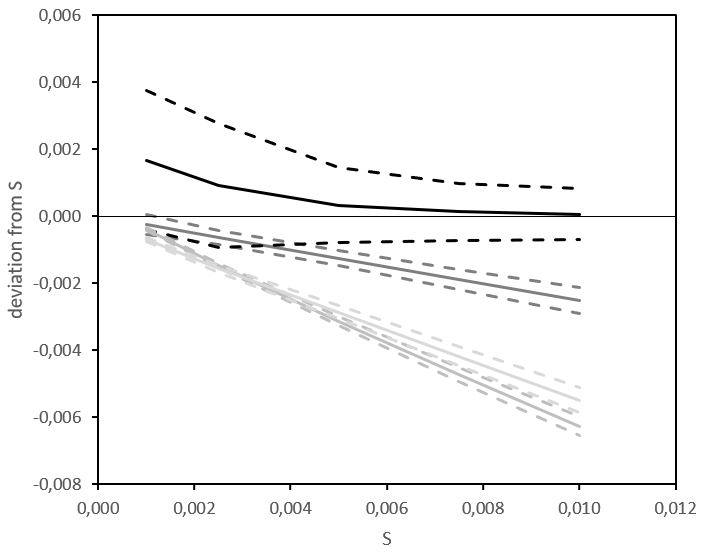}
\begin{minipage}{0.7\textwidth}\caption{Deviation of the estimators with respect to the true spread: the solid line is the bias, the dotted lines the bias $\pm$ the standard deviation of each estimator. The simulated dynamic is model 1 (left) or model 3 (right). The curves correspond, from the darker to the lighter, to $\widehat S_{3,1}$ (right graph only), $\widehat S_{1,1}$, CS, AGK$_2$.}
	\label{fig:Deviation_1}
\end{minipage}
\end{figure}

For model 1, $\widehat S_{1,1}$ has a very low bias, whatever the spread, as one can see in Figure~\ref{fig:Deviation_1}. Its standard deviation is a bit large, compared to other estimators, but AGK$_2$ (respectively CS) overestimates (resp.  underestimates) the true spread, with a low standard deviation which would lead to overconfidently rejecting the true value of $S$. The larger the spread, the larger the absolute bias of CS and AGK$_2$. We can also note that, when one looks at the ordinates axis, the amplitude of the deviation is small, compared to the other generating models. 

\begin{figure}[htbp]
	\centering
		\includegraphics[width=0.45\textwidth]{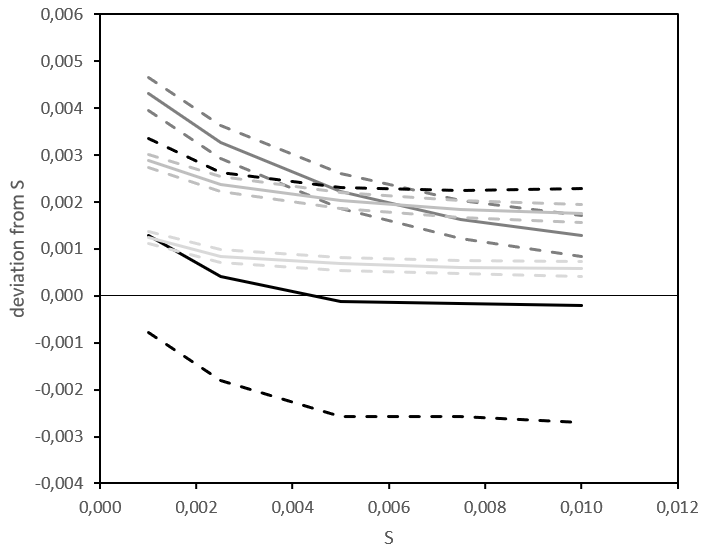} 
		\includegraphics[width=0.45\textwidth]{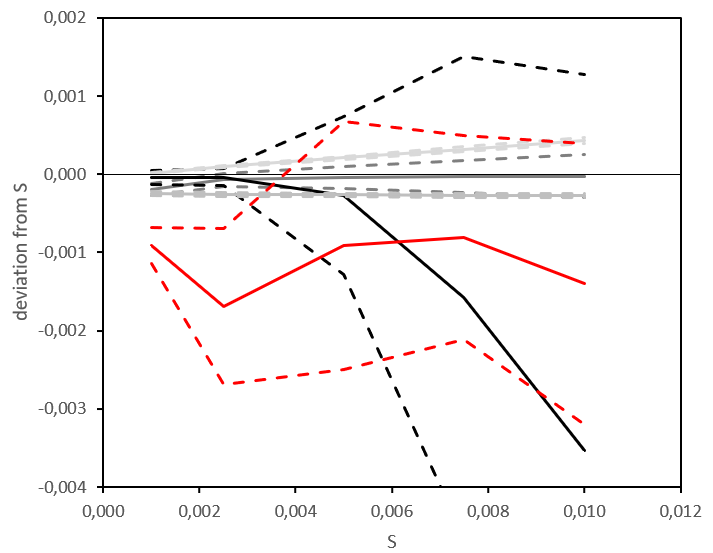}
\begin{minipage}{0.7\textwidth}\caption{Deviation of the estimators with respect to the true spread: the solid line is the bias, the dotted lines the bias $\pm$ the standard deviation of each estimator. The simulated dynamic is model 2, with $H=0.3$ (left) or $H=0.7$ (right). The curves correspond, from the darker to the lighter, to $\widehat S_{2,1}$, $\widehat S_{1,1}$, CS, AGK$_2$. The curves in red correspond to $\widehat S_{4,1}$.}
	\label{fig:Deviation_2}
\end{minipage}
\end{figure}

The results for model 2 are gathered in Figure~\ref{fig:Deviation_2}. When $H=0.3$, $\widehat S_{2,1}$ has a small bias. Its standard deviation is large, compared to that of $\widehat S_{1,1}$, CS, and AGK$_2$. These three other estimators however largely overestimate the true spread, with a narrow standard deviation that wrongly makes the true spread implausible. We also note that, whatever the model, the larger the spread, the lower the bias.

When $H=0.7$, $\widehat S_{2,1}$ tends to largely deviate from $S$ when $S$ increases. It thus seems to be relevant only for $S\leq 0.5\%$. The estimator $\widehat S_{4,1}$, which shows some erratic behaviour, caused by the numerical optimization, has the advantage to keep a limited deviation from $S$ when $S$ increases.

Last, when the data are generated by model 3, $\widehat S_{2,1}$ is the only of the four evaluated estimators to have the true spread in a reasonable confidence interval around the average estimate, as one can see in Figure~\ref{fig:Deviation_1}. Its bias also decreases for larger spreads, whereas the deviation of the other estimators from the true spreads becomes more pronounced. This disadvantageous effect is even stronger for CS and AGK$_2$ than for $\widehat S_{1,1}$, which is not supposed to handle correctly serial dependence.

\subsection{Simulations for infrequent trading}\label{sec:infreq}

For simulating illiquidity in the market, namely infrequent trading, we consider that the price at time $t$, which is observed every second as in Section~\ref{sec:resSimul}, is the same as at the previous time step $t-\tau/60$ with a probability $1-\pi^{\text{liqu}}$, because of an absence of trading event. The continuous trading simulated in Section~\ref{sec:resSimul} corresponds to $\pi^{\text{liqu}}=1$. For this study, we focus on the generating model 1.

\begin{figure}[htbp]
	\centering
		\includegraphics[width=0.7\textwidth]{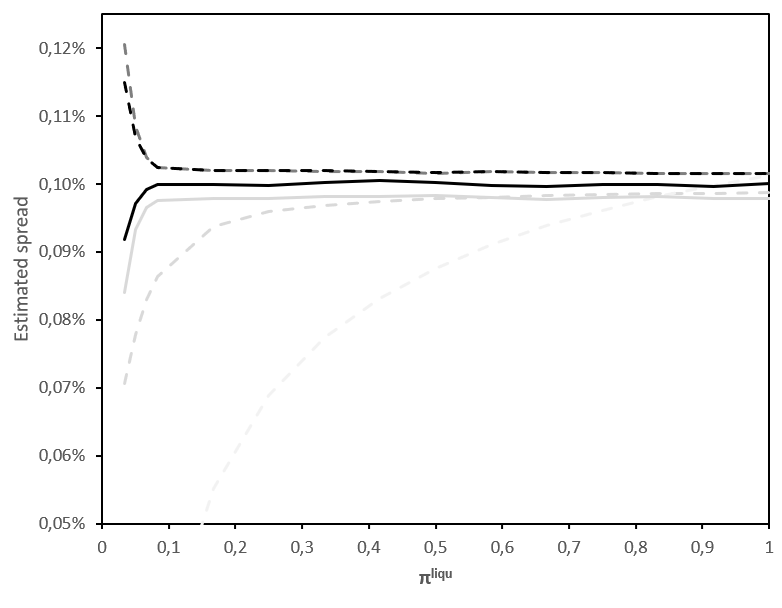} 
\begin{minipage}{0.7\textwidth}\caption{For a spread $S=0.1\%$ and a probability $\pi^{\text{liqu}}$ of observation of the price in the grid of step $\tau/60$, estimates using high-low ranges (dotted lines, from the lighter to the darker: CS, AR, AGK$_1$, AGK$_2$) and estimates using only one-minute close prices (solid lines, from the lighter to the darker: Roll, $\widehat S_{1,1}$).}
	\label{fig:Infreq}
\end{minipage}
\end{figure}

The results are displayed in Figure~\ref{fig:Infreq}. We observe that some estimators that use high and low prices are strongly impacted by infrequent trading. CS estimator behaves particularly poorly in this situation and requires continuous trading to lead to satisfying results. AR estimator, which mixes high and low with close prices, is less affected by illiquidity but it requires $\pi^{\text{liqu}}$ to be greater than 0.4, to limit the estimation error. AGK$_1$ and AGK$_2$, though they also use high and low prices, contain an explicit correction term for infrequent trading and thus behave satisfyingly in this situation.

Only very low probabilities $\pi^{\text{liqu}}$ of trading events, sampled every second, are able to bias the close price sampled at the lower one-minute frequency. Consequently, Roll and $\widehat S_{1,1}$ estimators are quite robust to illiquidity. These two estimators lead to satisfying results even for $\pi^{\text{liqu}}=0.1$, and, among all the six estimators displayed in Figure~\ref{fig:Infreq}, $\widehat S_{1,1}$ is the closest to the true spread, whatever $\pi^{\text{liqu}}$. This experiment illustrates the benefit of using estimators that do not use high and low prices when the market is illiquid.

\subsection{Application to real data}

For the real-life application, we consider the 40 constituents of the French stock index CAC 40. Our dataset includes open, high, low, and close prices in one-minute intervals, as well as bid-ask spreads sampled every five minutes. All the data are in the interval between the 19th September 2022 and the 31st March 2023. Using the intraday transaction prices, we estimate the daily spread using the standard estimators introduced in Section~\ref{sec:altEstim} and the new estimators introduced in Section~\ref{sec:estimators}. We compare these estimates with the true bid-ask spread, which is approximated as the daily average of the intraday bid-ask spreads. 

We assess the performance of each estimator using either the root mean square error (RMSE) or the mean absolute percentage error (MAPE). For an estimator $\widehat Z_{i,j}$ of the spread $S_{i,j}$, for a given day $i\in\llbracket 1,N_{\text{day}}\rrbracket$ and stock $j\in\llbracket 1,40\rrbracket$, these quantities are computed as
$$\text{RMSE}_j=\sqrt{\frac{1}{N_{\text{day}}}\sum_{i=1}^{N_{\text{day}}}\left[\log\left(\widehat Z_{i,j}\right)-\log\left(S_{i,j}\right)\right]^2}$$ 
and
$$\text{MAPE}_j=\frac{1}{N_{\text{day}}}\sum_{i=1}^{N_{\text{day}}}\left|\frac{\log\left(\widehat Z_{i,j}\right)-\log\left(S_{i,j}\right)}{\log\left(S_{i,j}\right)}\right|.$$
We note that we calculate the RMSE and the MAPE not for the spread itself but for its logarithm. This approach has already been used to avoid overweighting small (repectively large) spreads in MAPE (resp. RMSE)~\cite{AGK}.

Table~\ref{tab:CacAverage} displays the average errors over all the stocks. Three estimators have an average MAPE lower than 10\%, including $\widehat S_{1,1}$, the best in average for all the evaluated estimators, and $\widehat S_{2,1}$, the second best. A challenge in this application, compared to the simulation study, is that we don't know if a potential serial dependence stems from prices or from microstructure noise. Therefore, deciding to use $\widehat S_{2,1}$ or $\widehat S_{3,1}$ is a difficult task without a deeper preliminary empirical study of the dataset. We also have excluded $\widehat S_{4,1}$ from this study, since simulations showed it was less relevant than $\widehat S_{2,1}$ and $\widehat S_{3,1}$.

\begin{table}[htbp]
\centering
\begin{tabular}{|l|c|c|}
\hline
Estimator & Average RMSE & Average MAPE \\
\hline
$\widehat S_{1,1}$ & 0.613 & 6.06\% \\
$\widehat S_{2,1}$ & 0.685 & 7.01\% \\
Roll & 0.725 & 7.01\% \\
AR & 1.049 & 11.47\% \\
AGK$_1$ & 1.158 & 12.69\% \\
AGK$_2$ & 1.158 & 12.76\% \\
$\widehat S_{3,1}$ & 1.321 & 14.88\% \\
CS & 1.629 & 20.21\% \\
\hline
\end{tabular}
\begin{minipage}{0.9\textwidth}\caption{Average RMSE and MAPE over the 40 constituents of the CAC 40 index, for various spread estimators.}
\label{tab:CacAverage}
\end{minipage}
\end{table}

Though results are encouraging for $\widehat S_{2,1}$ and $\widehat S_{3,1}$, the simple estimator $\widehat S_{1,1}$, which does not correct serial dependence, is even better. We however believe that $\widehat S_{2,1}$ and $\widehat S_{3,1}$ may be more relevant than  $\widehat S_{1,1}$ in some practical frameworks. Indeed, beyond the aggregated view, $\widehat S_{2,1}$ is for example the best estimator for 8 single stocks, according to RMSE (same result for MAPE). More importantly, we have considered very liquid stocks at a time scale at which it is difficult to observe serial dependence. We prepare another article presenting a detailed empirical evaluation of these estimators using data at a much higher frequency.

We also observe that the performance of the estimator depends on the total market capitalization for each stock. For a stock $j$, we define $R^{\text{capi}}_j$ as the rank of the market capitalization for stock $j$ among the 40 constituents of the CAC 40. The lowest capitalization has the rank 1. We also define $R^{\text{RMSE}}_{j,k}$ (respectively $R^{\text{MAPE}}_{j,k}$), with $k\in\llbracket 1,8\rrbracket$, as the rank of RMSE$_j$ (resp. MAPE$_j$) for estimator $k$ among all the 8 estimators. The lowest error has the rank 1. We quantify the dependence between the performance of an estimator and the capitalization using a nonlinear approach, namely Spearman's rank correlation: Spearman$_{\text{RMSE}}$ (respectively Spearman$_{\text{MAPE}}$) is the correlation between $R^{\text{capi}}_j$ and $R^{\text{RMSE}}_{j,k}$ (resp. $R^{\text{MAPE}}_{j,k}$) over $j\in\llbracket 1,40\rrbracket$. Results are gathered in Table~\ref{tab:CacSpearman}. Positive (respectively negative) values indicate that the corresponding estimator is relatively more accurate for small (resp. large) capitalizations. The estimators AGK$_1$ and to a lesser extent $\widehat S_{1,1}$ seem to be a good choice whatever the capitalization, but $\widehat S_{2,1}$, $\widehat S_{3,1}$, and AGK$_2$ improve their relative accuracy for smaller capitalizations, whereas we observe the opposite for Roll, AR, and CS.

\begin{table}[htbp]
\centering
\begin{tabular}{|l|c|c|}
\hline
Estimator & Spearman$_{\text{RMSE}}$ & Spearman$_{\text{MAPE}}$ \\
\hline
$\widehat S_{1,1}$ & 0.09 & 0.21 \\
$\widehat S_{2,1}$ & 0.54 & 0.55 \\
Roll & -0.30 & -0.42 \\
AR & -0.47 & -0.55 \\
AGK$_1$ & -0.03 & 0.03 \\
AGK$_2$ & 0.31 & 0.23 \\
$\widehat S_{3,1}$ & 0.53 & 0.53 \\
CS & -0.59 & -0.59 \\
\hline
\end{tabular}
\begin{minipage}{0.9\textwidth}\caption{Spearman's rank correlation coefficient between the market capitalization and either the RMSE or the MAPE over the 40 constituents of the CAC 40 index and the 8 spread estimators.}
\label{tab:CacSpearman}
\end{minipage}
\end{table}

\begin{figure}[htbp]
	\centering
		\includegraphics[width=0.45\textwidth]{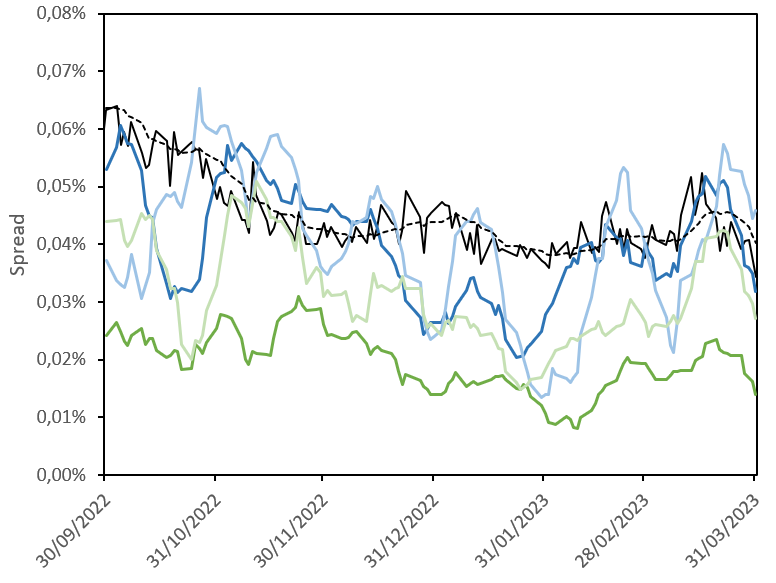} 
		\includegraphics[width=0.45\textwidth]{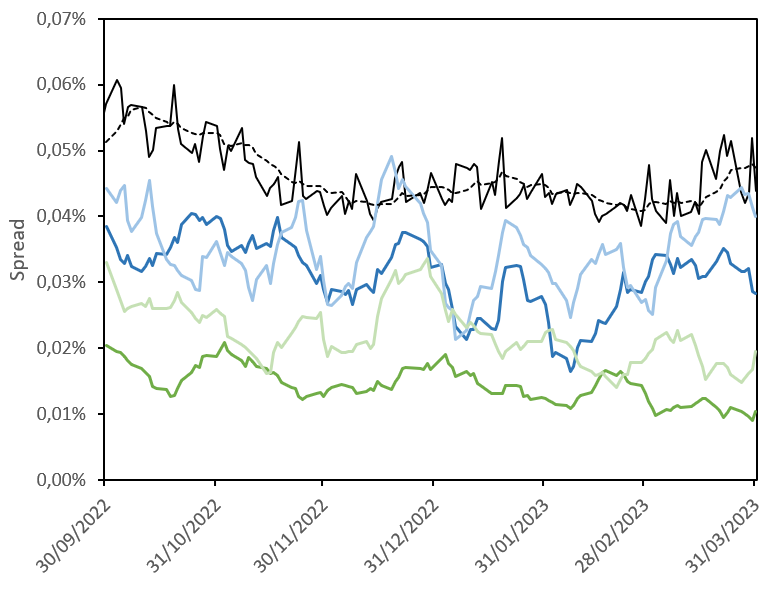} 
\begin{minipage}{0.7\textwidth}\caption{Evolution of the spread for the stocks RMS (left) and EN (right). In black is the true spread (solid line) and its 10-day average (dotted line). The other curves are 10-day averages of the following estimators: $\widehat S_{1,1}$ (dark blue), $\widehat S_{2,1}$ (light blue), Roll (light green), and AR (dark green).}
	\label{fig:spread_singlestocks}
\end{minipage}
\end{figure}

Finally, Figure~\ref{fig:spread_singlestocks} displays the evolution during 6 month of the estimated spreads for two different stocks. We focus on the stock RMS (Herm\`es), which is one of the highest capitalizations of the CAC 40 index, and on the stock EN (Bouygues), one of the lowest capitalizations in this index. We only represent the four more relevant estimators according to Table~\ref{tab:CacAverage}, that is $\widehat S_{1,1}$, $\widehat S_{2,1}$, Roll, and AR. These four estimators are quite erratic. We thus represent the average during the last 10 days of these estimators. We see from this graph that AR and, to a lesser extent, Roll estimators are significantly lower than the two others in the case of EN, while we cannot disqualify Roll's estimator so obviously in the case of RMS. It thus confirms that other spread estimators than the very classical ones are useful, in particular for low-capitalization stocks.

\section{Conclusion}\label{sec:conclusion}

We have proposed four kinds of estimators of bid-ask spreads, based on variances of increments of observed prices, at several time scales. The first kind of estimator is suitable in the absence of serial dependence. The other estimators are adapted to serial dependence either of the price returns, with a model based on an fBm, or of the trades, with a binarized Ornstein-Uhlenbeck process. Beyond the theoretical properties of these estimators, such as the consistency and asymptotic Gaussianity, simulations show the good performance of each of these estimators in the context for which it has been built, compared to classical bid-ask spread estimators. An application to real data with liquid stocks, observed at a low frequency, shows that estimators intended to filter serial dependence are more accurate, with respect to more classical estimators, when the market capitalization is smaller. This stirs up the intuition that one should favour these estimators in a less liquid environment. Future developments could consist in adapting bid-ask spread estimators based on the high-low range to serial dependence. Moreover, in the perspective of building time series of bid-ask spreads at a high frequency, a natural extension of our work would consist in considering also a dependence between trades and price increments, as suggested by tick-by-tick data~\cite{RR,HLR}.

\bibliographystyle{plain}
\bibliography{biblioEstimSpread}

\appendix

\section{Useful lemmas}\label{sec:lemmas}

We are first interested in discrete random variables $Y_1,...,Y_n$ following a two-point distribution with values in $\{-1,1\}$. If $\proba(Y_1=1)=p$, we note $Y\sim\mathcal D_{\{-1,1\}}(p)$. In the case where $p=1/2$, this distribution is simply the Rademacher distribution, in which case $\E[Y_1]=0$ and $\Var(Y_1)=1$. The aim of the following lemma is to decompose a set of correlated Rademacher variables in independent two-point variables. The correlation structure allowed by the lemma encompasses independent variables, like in Sections~\ref{sec:zeroCorrel} and~\ref{sec:fBm}, as well as time series with an exponentially decaying correlation, like in Sections~\ref{sec:correlTrades} and~\ref{sec:fBmAndCorrelTrades}.

\begin{lem}\label{lem:decompoIndep}
Let $Y_1,...,Y_n$ be $n$ variables in $\mathcal D_{\{-1,1\}}(1/2)$ with correlation $\corr(Y_i,Y_j)=\rho_{i,j}$, such that $\rho_{k,i}=\rho_{k,j}\rho_{j,i}$ for $i\leq j\leq k$. Let's introduce $n-1$ variables $Z_i\sim\mathcal D_{\{-1,1\}}((1+\rho_{i,i-1})/2)$, for $i\in\llbracket 2,n\rrbracket$, such that the $n$ variables $Y_1,Z_2,...,Z_n$ are mutually independent. Then, the vectors $(Y_1,...,Y_n)^T$ and $(Y_1,Y_1Z_2,...,Y_1\prod_{j=2}^{n}Z_j)^T$ have the same distribution.
\end{lem}

Before proving Lemma~\ref{lem:decompoIndep}, it is worth noting that the variable $Z_i$ has the following moments: $\E[Z_i^{2k+1}]=\rho_{i,i-1}$ and $\E[Z_i^{2k}]=1$, for $k\in\mathbb Z$.

\begin{proof}
Let $Y_i'=Y_1\prod_{j=2}^{i}Z_j$. We have $Y_i\sim\mathcal D_{\{-1,1\}}(1/2)$ and $Y'_i\sim\mathcal D_{\{-1,1\}}(p)$, with $p$ to be determined. The expected value of a variable in $\mathcal D_{\{-1,1\}}(p)$ is $2p-1$. By independence, we have $\E[Y_1\prod_{j=2}^{i}Z_j]=\E[Y_1]\prod_{j=2}^{i}\E[Z_j]=0$, so that $p=1/2$. Therefore, $Y_i$ and $Y_i'$ have the same marginal distribution. We now focus on the dependence structure. Let $i\leq j$. Then, due to the particular moments of the distribution, the correlation between $Y_i'$ and $Y_j'$ is equal to $\E[Y_i'Y_j']$, which, using the independence, is equal to $\E[Y_1^2\prod_{k=2}^{i}Z_k^2\prod_{k=i+1}^{j}Z_k]=\prod_{k=i+1}^{j}\E[Z_k]=\prod_{k=i+1}^{j}\rho_{k,k-1}$. This last product is equal to $\rho_{j,i}$ by assumption, which is precisely the correlation between $Y_i$ and $Y_j$.
\end{proof}

In the next lemma, we will use the decomposition of the $n$ variables $Y_1,...,Y_n$ in products of $n$ independent variables. This lemma provides a cokurtosis of two random variables mixing correlated Gaussian variables and correlated two-point variables. It will be useful for determining the variance of the empirical variance of price log-returns in the four specifications studied in this paper.

\begin{lem}\label{lem:cokurtosis}
Let $G_a,G_b\sim\mathcal N(0,\sigma_G^2)$ of correlation $\rho_G$, $Y_1$, $Y_2$, $Y_3$ and $Y_4$ be four Rademacher variables independent of $G_a$ and $G_b$, with $\corr(Y_i,Y_j)=\rho_{i,j}$, such that $\rho_{k,i}=\rho_{k,j}\rho_{j,i}$ for $i\leq j\leq k$. We define
$$\left\{\begin{array}{ccl}
\mathcal K^- & = & \E\left[(G_a+Y_2-Y_1)^2(G_b+Y_4-Y_3)^2\right] \\
\mathcal K^+ & = & \E\left[(G_a+Y_3-Y_1)^2(G_b+Y_4-Y_2)^2\right].
\end{array}\right.$$
Then, we get
$$\begin{array}{ccl}
\mathcal K^- & = & (1+2\rho_G^2)\sigma_G^4 + 2\sigma_G^2(2-\rho_{4,3}-\rho_{2,1}) \\
 & & + 4(1-\rho_{2,1})(1-\rho_{4,3}) + 4\rho_G\sigma_G^2(\rho_{3,1}-\rho_{4,1}-\rho_{3,2}+\rho_{4,2})
\end{array}$$
and
$$\begin{array}{ccl}
\mathcal K^+ & = & (1+2\rho_G^2)\sigma_G^4 + 2\sigma_G^2(2-\rho_{4,2}-\rho_{3,1}) \\
 & & + 4(1-\rho_{4,2}-\rho_{3,1}+\rho_{4,3}\rho_{2,1}) + 4\rho_G\sigma_G^2(\rho_{4,3}-\rho_{3,2}-\rho_{4,1}+\rho_{2,1}).
\end{array}$$
\end{lem}

\begin{proof}
Using the Cholesky decomposition, we write $G_b=\rho_G G_a+\sqrt{1-\rho_G^2}G_{b,a}$, where $G_{b,a}\sim\mathcal N(0,\sigma_G^2)$ is independent of $G_a$. As a consequence,
$$\begin{array}{ccl}
\E\left[G_a^2G_b^2\right] & = & \rho_G^2\E\left[G_a^4\right]+2\rho_G\sqrt{1-\rho_G^2}\E\left[G_a^3\right]\E\left[G_{b,a}\right]+(1-\rho_G^2)\E\left[G_a^2\right]\E\left[G_{b,a}^2\right] \\
 & = & (1+2\rho_G^2)\sigma_G^4,
\end{array}$$
where we used the Gaussian assumption, which leads to a kurtosis of 3. We also have, for $j\geq i$, using the decomposition in a product of independent variables, as in Lemma~\ref{lem:decompoIndep}, and the assumption on the product of correlations,
$$\begin{array}{ccl}
\E\left[(Y_{i}-Y_j)^2\right] & = & \E\left[Y_{i}^2\left(1-\prod_{k=i+1}^{j}Z_k\right)^2\right] \\
 & = & \E\left[Y_{i}^2\right]\left(1-2\prod_{k=i+1}^{j}\E\left[Z_k\right]+\prod_{k=i+1}^{j}\E\left[Z_k^2\right]\right) \\
 & = & 2-2\prod_{k=i+1}^{j}\rho_{k,k-1} \\
 & = & 2(1-\rho_{j,i}), 
\end{array}$$
as well as, using a standard expansion of the product, 
$$\left\{\begin{array}{ccl}
\E[(Y_2-Y_1)(Y_4-Y_3)] & = & \rho_{3,1}-\rho_{4,1}-\rho_{3,2}+\rho_{4,2} \\
\E[(Y_3-Y_1)(Y_4-Y_2)] & = & \rho_{4,3}-\rho_{3,2}-\rho_{4,1}+\rho_{2,1}
\end{array}\right.$$
and
$$\begin{array}{ccl}
\E\left[(Y_2-Y_1)^2(Y_4-Y_3)^2\right] & = & \E\left[Y_1^4(1-Z_2)^2(1-Z_4)^2Z_3^2Z_2^2\right] \\
 & = & \E\left[Y_1^4\right]\E\left[(1-Z_2)^2Z_2^2\right]\E\left[(1-Z_4)^2\right]\E\left[Z_3^2\right] \\
 & = & \E\left[(1-Z_2)^2\right]\E\left[(1-Z_4)^2\right] \\
 & = & 4(1-\rho_{2,1})(1-\rho_{4,3})
\end{array}$$
and, similarly,
$$\begin{array}{ccl}
\E\left[(Y_3-Y_1)^2(Y_4-Y_2)^2\right] & = & \E\left[Y_1^4(Z_3Z_2-1)^2(Z_4Z_3-1)^2Z_2^2\right] \\
 & = & 4\E\left[(1-Z_3Z_2)(1-Z_4Z_3)\right] \\
 & = & 4\E\left[1-Z_4Z_3-Z_3Z_2+Z_4Z_3^2Z_2\right] \\
 & = & 4(1-\rho_{4,3}\rho_{3,2}-\rho_{3,2}\rho_{2,1}+\rho_{4,3}\rho_{2,1}) \\
 & = & 4(1-\rho_{4,2}-\rho_{3,1}+\rho_{4,3}\rho_{2,1}).
\end{array}$$
Using all these preliminary results, the zero mean and the independence between the Gaussian and the two-point variables, we get
$$\begin{array}{ccl}
\mathcal K^- & = & \E\left[G_a^2G_b^2\right]+\E\left[G_a^2\right]\E\left[(Y_4-Y_3)^2\right]+\E\left[G_b^2\right]\E\left[(Y_2-Y_1)^2\right] \\
 & & +\E\left[(Y_2-Y_1)^2(Y_4-Y_3)^2\right]+4\E\left[G_aG_b\right]\E\left[(Y_2-Y_1)(Y_4-Y_3)\right] \\
 & = & (1+2\rho_G^2)\sigma_G^4 + 2\sigma_G^2(2-\rho_{4,3}-\rho_{2,1}) \\
 & & + 4(1-\rho_{2,1})(1-\rho_{4,3}) + 4\rho_G\sigma_G^2(\rho_{3,1}-\rho_{4,1}-\rho_{3,2}+\rho_{4,2})
\end{array}$$
and
$$\begin{array}{ccl}
\mathcal K^+ & = & \E\left[G_a^2G_b^2\right]+\E\left[G_a^2\right]\E\left[(Y_4-Y_2)^2\right]+\E\left[G_b^2\right]\E\left[(Y_3-Y_1)^2\right] \\
 & & +\E\left[(Y_3-Y_1)^2(Y_4-Y_2)^2\right]+4\E\left[G_aG_b\right]\E\left[(Y_3-Y_1)(Y_4-Y_2)\right] \\
 & = & (1+2\rho_G^2)\sigma_G^4 + 2\sigma_G^2(2-\rho_{4,2}-\rho_{3,1}) \\
 & & + 4(1-\rho_{4,2}-\rho_{3,1}+\rho_{4,3}\rho_{2,1}) + 4\rho_G\sigma_G^2(\rho_{4,3}-\rho_{3,2}-\rho_{4,1}+\rho_{2,1}).
\end{array}$$
\end{proof}

\section{Proofs for the standard zero-autocorrelation market model}

\subsection{Proof of Proposition~\ref{pro:covIncrSq1}}\label{sec:covIncrSq1}

\begin{proof}
Thanks to Assumptions~\ref{assum:indepProcesses} and~\ref{assum:zeroCorrel}, $K(u,\delta)16/S^4$ is equal to the quantity $\mathcal K^+$ when $\delta> 0$ or $\mathcal K^-$ when $\delta\leq 0$, as displayed in Lemma~\ref{lem:cokurtosis}, with $\sigma_G^2=\frac{4u\sigma^2}{S^2}$ and with correlation parameters which depend on the value of $\delta$.

We start with the case $\delta< 0$, corresponding to strictly disjoint increments and thus to $\rho_G=0$ and, for all $i\neq j$, $\rho_{i,j}=0$. The corresponding cokurtosis of Lemma~\ref{lem:cokurtosis} is $\mathcal K^-=\sigma_G^4 + 4\sigma_G^2 + 4$, so that
$$\begin{array}{ccl}
K(u,\delta) & = & u^2\sigma^4+ u\sigma^2S^2 + S^4/4 \\
 & = & V(u)^2.
\end{array}$$ 

When $\delta=0$, increments do not overlap but have a bound in common, so that the only difference with the previous case is $\rho_{3,2}=1$. This difference has no impact on $\mathcal K^-$ and thus we still have $K(u,\delta)=V(u)^2$.

When $\delta\in(0,u)$, we have strictly overlapping increments and thus $\rho_G=\delta/u$ and, for all $i\neq j$, $\rho_{i,j}=0$. In this case, the relevant cokurtosis in Lemma~\ref{lem:cokurtosis} is $\mathcal K^+=(1+2\delta^2/u^2)\sigma_G^4 + 4\sigma_G^2 + 4$, so that 
$$\begin{array}{ccl}
K(u,\delta) & = & (u^2+2\delta^2)\sigma^4+ u\sigma^2S^2 + S^4/4 \\
 & = & V(u)^2+2\delta^2\sigma^4.
\end{array}$$ 

Finally, when $\delta=u$, that is when the increments are the same, we have $\rho_G=1$, $\rho_{2,1}=\rho_{4,3}=1$, and $\rho_{i,j}=0$ for other values of $(i,j)$ with $i\neq j$. Then, $\mathcal K^+=3\sigma_G^4 + 12\sigma_G^2+8$ and
$$\begin{array}{ccl}
K(u,\delta) & = & 3u^2\sigma^4+ 3u\sigma^2S^2 + S^4/2 \\
 & = & V(u)^2+2u^2\sigma^4+2u\sigma^2S^2+S^4/4.
\end{array}$$ 
\end{proof}

\subsection{Proof of Proposition~\ref{pro:momentV1}}\label{sec:momentV1}

\begin{proof}
The result on the first moment is a straightforward consequence of the linearity of the expectation and of equation~\eqref{eq:varianceIncrPrix}.

For $v=3$, using the stationarity of the increments, the second moment is
\begin{equation}\label{eq:momentV1_preuve1}
\E\left[\widehat V_3(n,L)^2\right]=\frac{1}{k_3(n,L)^2}\sum_{i=0}^{k_3(n,L)-1}\left\{K(L\tau,L\tau)+2\sum_{j=i+1}^{k_3(n,L)-1}K(L\tau,(i-j+(i-j+1)L)\tau)\right\}.
\end{equation}
All the terms in each sum are equal, so that we just have to count the number of terms in each:
$$\begin{array}{ccl}
\E\left[\widehat V_3(n,L)^2\right] & = & \frac{k_3(n,L)K(L\tau,L\tau)+k_3(n,L)(k_3(n,L)-1)K(L\tau,0)}{k_3(n,L)^2} \\
 & = & V(L\tau)^2 + \frac{2\sigma^4L^2\tau^2 + 2\sigma^2L\tau S^2 + \frac{S^4}{4}}{k_3(n,L)},
\end{array}$$
which, by subtracting $V(L\tau)^2$, leads to the variance.

For $v=2$, we have the same kind of double sum as in equation~\eqref{eq:momentV1_preuve1}, except that $k_3(n,L)$ is to be replaced by $k_2(n,L)$:
\begin{equation}\label{eq:momentV1_preuve1b}
\E\left[\widehat V_2(n,L)^2\right]=\frac{1}{k_2(n,L)^2}\sum_{i=0}^{k_2(n,L)-1}\left\{K(L\tau,L\tau)+2\sum_{j=i+1}^{k_2(n,L)-1}K(L\tau,(i-j+1)L\tau)\right\},
\end{equation}
with the terms $K(L\tau,(i-j+1)L\tau)$ all equal to $K(L\tau,0)$. We thus get a very similar result as in the case $v=3$:
$$\Var\left[\widehat V_2(n,L)\right]=\frac{2\sigma^4L^2\tau^2 + 2\sigma^2L\tau S^2 + \frac{S^4}{4}}{k_2(n,L)}.$$

For $v=1$, the same kind of decomposition as in equation~\eqref{eq:momentV1_preuve1} leads to $k_1(n,L)^2$ terms, $k_1(n,L)$ of which corresponding to synchronized increments and thus to $K(L\tau,L\tau)$, $2(k_1(n,L)-l)$ terms corresponding to overlapping increments with a duration of overlap being equal to $(L-l)\tau$, all the other terms corresponding to non-overlapping increments and thus to $K(L\tau,0)$:
\begin{equation}\label{eq:momentV1_preuve1c}
\begin{array}{ccl}
\E\left[\widehat V_1(n,L)^2\right] & = & \frac{1}{k_1(n,L)^2}\left(k_1(n,L)K(L\tau,L\tau) + \sum_{l=1}^{L-1}2(k_1(n,L)-l)K(L\tau,(L-l)\tau) \right. \\
 & & \left.+ \left(k_1(n,L)^2-k_1(n,L)-\sum_{l=1}^{L-1}2(k_1(n,L)-l)\right)K(L\tau,0)\right) \\
 & = & V(L\tau)^2 + \frac{1}{k_1(n,L)^2}\left(k_1(n,L)\left(2\sigma^4L^2\tau^2+2\sigma^2L\tau S^2+\frac{S^4}{4}\right) \right. \\
 & & \left. + 4\sigma^4\sum_{m=1}^{L-1}(k_1(n,L)-L+m)m^2\tau^2\right) \\
 & = & V(L\tau)^2 + \frac{1}{k_1(n,L)}\left(2\sigma^4L^2\tau^2+2\sigma^2L\tau S^2+\frac{S^4}{4} + 4\sigma^4\tau^2\sum_{m=1}^{L-1}m^2\right) \\
 & & + \mathcal{O}\left(\frac{1}{k_1(n,L)^2}\right) \\
 & = & V(L\tau)^2 + \frac{1}{k_1(n,L)}\left(\frac{2}{3}\sigma^4L(1+2L^2)\tau^2+2\sigma^2L\tau S^2+\frac{S^4}{4}\right) + \mathcal{O}\left(\frac{1}{k_1(n,L)^2}\right).
\end{array}
\end{equation}
Noting that $k_1(n,L)=n-L$, we get the expected result.
\end{proof}

\subsection{Proof of Theorem~\ref{thm:S1}}\label{sec:S1}

\begin{proof}
The absence of bias is a direct consequence of Proposition~\ref{pro:momentV1} and equation~\eqref{eq:S_vs_moments1}.

The increments of $p$ in $\widehat V_3(n,L)$ being independent of each other, $\widehat V_3(n,L)$ converges almost surely toward $V(L\tau)$, by the strong law of large numbers. We obtain a similar result for $\widehat V_3(n,L')$ toward $V(L'\tau)$. The continuous mapping theorem applied with $g:(x,y)\mapsto 2(L'x-Ly)/(L'-L)$ thus leads to the almost sure convergence of $\widehat{S^2_{1,3}}(n,L,L')$ toward $2(L'V(L\tau)-LV(L'\tau))/(L'-L)$, which is equal to $S^2$, according to equation~\eqref{eq:S_vs_moments1}.

Regarding the convergence of $\widehat{S^2_{1,v}}(n,L,L')$ for $v\in\{1,2\}$, we first note, using Proposition~\ref{pro:momentV1}, that $\widehat V_v(n,L)$ converges in probability toward $V(L\tau)$ because its asymptotic quadratic risk is zero. Then, using also the continuous mapping theorem, one concludes that $\widehat{S^2_{1,v}}(n,L,L')$ converges in probability toward $S^2$.

For the convergence in distribution, we first note that $\widehat V_1(n,L)$ (respectively $\widehat V_2(n,L)$, $\widehat V_3(n,L)$) is a sum of $L$-dependent (resp. $1$-dependent, $0$-dependent) identically distributed variables. Therefore, following the formalism of \cite[Theorem 27.4]{Billingsley} whatever $v\in\{1,2,3\}$, $\widehat V_v(n,L)$ is trivially a sum of an $\alpha$-mixing sequence, with $\alpha_n=0$ for $n$ sufficiently large, so that the central limit theorem applies: 
$$\sqrt{n}\left(\widehat V_v(n,L)-V(L\tau)\right)\overset{d}{\longrightarrow} \mathcal N\left(0,\zeta_v(L)\sigma_{1,v}^2(L)\right).$$
Its multivariate version, applied to the vector $(\widehat V_v(n,L),\widehat V_v(n,L'))^T$, also holds, because this vector is a sum of vectors which are also dependent at a finite range only and thus $\alpha$-mixing (the case $v=3$ and $L'=m(L+1)-1$, presented in the last part of the proof, shows how the sum of vectors can be decomposed in such a way as to get rid of the dependence):
\begin{equation}\label{eq:TCL_dim2}
\sqrt{n}\left(\left(\begin{array}{c}
\widehat V_v(n,L) \\
\widehat V_v(n,L')
\end{array}\right) - \left(\begin{array}{c}
V(L\tau) \\
V(L'\tau)
\end{array}\right)\right)\overset{d}{\longrightarrow} \mathcal N_2\left(0,\Sigma\right),
\end{equation}
where $\mathcal N_2$ is the bivariate Gaussian distribution and 
$$\Sigma=\left(\begin{array}{cc}
\zeta_v(L)\sigma_{1,v}^2(L) & \sqrt{\zeta_v(L)\zeta_v(L')}\sigma_{1,v}(L)\sigma_{1,v}(L')r_{1,v}(L,L') \\
\sqrt{\zeta_v(L)\zeta_v(L')}\sigma_{1,v}(L)\sigma_{1,v}(L')r_{1,v}(L,L') & \zeta_v(L')\sigma_{1,v}^2(L')
\end{array}\right).$$ 
Then, we want to apply the delta method with the function $g$. The gradient of $g$ is
$$\nabla g(x,y)=\frac{2}{L'-L}\left(\begin{array}{c}
L' \\
-L
\end{array}\right).$$
We also have $g\left(\widehat V_v(n,L),\widehat V_v(n,L')\right)=\widehat{S^2_{1,v}}(n,L,L')$ and, after equation~\eqref{eq:S_vs_moments1}, $g\left(V(L\tau),V(L'\tau)\right)=S^2$. Therefore,
$$\sqrt{n}\left(\widehat{S^2_{1,v}}(n,L,L') - S^2\right)\overset{d}{\longrightarrow} \mathcal N\left(0,\gamma_{1,v}(L,L')\right),$$
with
$$\begin{array}{ccl}
\gamma_{1,v}(L,L') & = & \frac{4}{(L'-L)^2}\left(\begin{array}{c}
L' \\
-L
\end{array}\right)^T\Sigma\left(\begin{array}{c}
L' \\
-L
\end{array}\right) \\
 & = & \frac{4}{(L'-L)^2} \left(L'^2\zeta_v(L)\sigma_{1,v}^2(L) + L^2\zeta_v(L')\sigma_{1,v}^2(L') - 2LL'\sqrt{\zeta_v(L)\zeta_v(L')}\sigma_{1,v}(L)\sigma_{1,v}(L')r_{1,v}(L,L')\right).
\end{array}$$
This proves equation~\eqref{eq:TCL1}.

We now focus on the case $v=3$, $L'=m(L+1)-1$. We start by applying the central limit theorem to the vector 
\begin{equation}\label{eq:thm_S1_vectSomme}
\left(\begin{array}{c}
\widehat V_3(n,L) \\
\widehat V_3(n,m(L+1)-1)
\end{array}\right)
=\frac{M}{n}\sum_{i=0}^{(n/M)-1}\left(\begin{array}{c}
\frac{1}{m}\sum_{j=0}^{m-1}\left(p_{(j+iM+(j+1)L)\tau}-p_{(j+iM+jL)\tau}\right)^2 \\
\left(p_{(i+(i+1)(M-1))\tau}-p_{(i+i(M-1))\tau}\right)^2
\end{array}\right),
\end{equation}
in which, for the convenience of the notation and without any consequence on the result of the central limit theorem, we assume that $n$ is a multiple of $M$, where we note $M=m(L+1)$ the number of time steps between two starts of consecutive low-frequency increments in $\widehat V_3(n,m(L+1)-1)$. In the above vector, we have split the $n/(L+1)$ disjoint increments appearing in $\widehat V_3(n,L)$ in $n/M$ groups of $m$ consecutive disjoint increments, so that each group overlaps with a single increment appearing in $\widehat V_3(n,m(L+1)-1)$. More precisely, noting $A_{i,j}=p_{(j+iM+(j+1)L)\tau}-p_{(j+iM+jL)\tau}$ and $B_{i,j}=p_{(j+1+iM+(j+1)L)\tau}-p_{(j+iM+(j+1)L)\tau}$, a low-frequency increment is the sum of the successive non-overlapping increments, with one bound in common, $A_{i,0}$, $B_{i,0}$, $A_{i,1}$, $B_{i,1}$, ..., $B_{i,m-2}$, $A_{i,m-1}$. We are thus interested in the vector 
$$\textbf{Y}_i=\left(\begin{array}{c}
 \frac{1}{m}\sum_{j=0}^{m-1} A_{i,j}^2 \\
 \left(\sum_{j=0}^{m-1} A_{i,j} + \sum_{j=0}^{m-2} B_{i,j}\right)^2
\end{array}\right),$$
for $i\in\llbracket 0,(n/M)-1\rrbracket$. The vector defined in equation~\eqref{eq:thm_S1_vectSomme} is the sum of all the $\textbf{Y}_i$, which are all independent to each other and identically distributed. We note $\Sigma^{Y}$ the covariance matrix of a vector $\textbf{Y}_i$. We are looking for an expression of the covariance $\Sigma^{Y}_{12}$ between the two components of $\textbf{Y}_i$:
$$\Sigma^{Y}_{12} = \Sigma^{Y}_{21} = \Cov\left(\frac{1}{m}\sum_{j=0}^{m-1}A_{i,j}^2,\left(\sum_{j'=0}^{m-1}A_{i,j'}+\sum_{j'=0}^{m-2}B_{i,j'}\right)^2\right).$$
Since $A_{i,j}$ is independent from $A_{i,j'}$ for $j'\neq j$ and from $B_{i,j'}$ for $j'\notin\{j-1,j\}$, since $\E(A_{i,j})=\E(B_{i,j})=0$, and since the $A_{i,j}$ (respectively the $B_{i,j}$) are identically distributed, the covariance can be simplified to the following expression, for any $i$ and any $j$ such that the variables are defined,
$$\begin{array}{ccl}
\Sigma^{Y}_{12} & = & \Cov(A_{i,j}^2,A_{i,j}^2) + \frac{m-1}{m}\Cov(A_{i,j}^2,B_{i,j}^2) + \frac{m-1}{m}\Cov(A_{i,j}^2,B_{i,j-1}^2) \\
 & & + 2\frac{m-2}{m}\Cov(A_{i,j}^2,B_{i,j-1}B_{i,j}) + 2\frac{m-1}{m}\Cov(A_{i,j}^2,A_{i,j}B_{i,j}) + 2\frac{m-1}{m}\Cov(A_{i,j}^2,A_{i,j}B_{i,j-1}) \\
 & & + 2\frac{m-1}{m}\Cov(A_{i,j}^2,B_{i,j}A_{i,j+1}) + 2\frac{m-1}{m}\Cov(A_{i,j}^2,B_{i,j-1}A_{i,j-1}),
\end{array}$$
where
$$\Cov(A_{i,j}^2,A_{i,j}^2)=K(L\tau,L\tau)-V(L\tau)^2,$$
whose expression is provided in Proposition~\ref{pro:covIncrSq1},
$$\Cov(A_{i,j}^2,B_{i,j}^2)=\Cov(A_{i,j}^2,B_{i,j-1}^2)=\Cov(\eta_.^2,\eta_.^2)=0,$$
noting that the square of the random variable $\eta.$, which has a value in $\{-S/2,S/2\}$, is a constant (equal to $S^2/4$),
$$\begin{array}{ccl}
\Cov(A_{i,j}^2,B_{i,j-1}B_{i,j}) & = & \Cov(\eta_.^2+\eta_{.+L\tau}^2-2\eta_{.}\eta_{.+L\tau},-\eta_{.}\eta_{.+L\tau}) \\
 & = & 2\Var(\eta_{.}\eta_{.+L\tau}) \\
 & = & 2\E(\eta_{.}^2)^2-2\E(\eta_{.})^4=\frac{S^4}{8},
\end{array}$$
because $\E(\eta_{.})=0$ and $\eta_{.}$ and $\eta_{.+L\tau}$ are independent and identically distributed,
$$\begin{array}{ccl}
\Cov(A_{i,j}^2,B_{i,j}A_{i,j+1}) & = & \Cov(A_{i,j}^2,B_{i,j-1}A_{i,j-1}) \\
 & = & \Cov(\eta_.^2+\eta_{.+L\tau}^2-2\eta_{.}\eta_{.+L\tau},(-\eta_{.+L\tau}+\eta_{.+(L+1)\tau})\eta_{.+(L+1)\tau})=0,
 \end{array}$$
for similar reasons, and, noting $A_{i,j}^{\star}=p^{\star}_{(j+iM+(j+1)L)\tau}-p^{\star}_{(j+iM+jL)\tau}$,
$$\begin{array}{ccl}
\Cov(A_{i,j}^2,A_{i,j}B_{i,j}) & = & \Cov(A_{i,j}^2,A_{i,j}B_{i,j-1}) \\
 & = & \Cov((A_{i,j}^{\star}-\eta_.+\eta_{.+L\tau})^2,-(A_{i,j}^{\star}-\eta_.+\eta_{.+L\tau})\eta_{.+L\tau}) \\
 & = & -2\Var(A_{i,j}^{\star}\eta_{.+L\tau})-2\Var(\eta_.\eta_{.+L\tau}) \\
 & = & -2\E((A_{i,j}^{\star})^2)\E(\eta_{.+L\tau}^2)-2\E(\eta_{.}^2)\E(\eta_{.+L\tau}^2) \\
 & = & -\sigma^2 L\tau\frac{S^2}{2}-\frac{S^4}{8} = -\frac{S^2}{2}V(L\tau)+\frac{S^4}{8}.
\end{array}$$
Finally, with all these intermediary calculations, we get:
\begin{equation}\label{eq:thm_S1_sigma12}
\begin{array}{ccl}
\Sigma^{Y}_{12} & = & K(L\tau,L\tau)-V(L\tau)^2+\frac{m-2}{m}\frac{S^4}{4}+4\frac{m-1}{m}\left(-\frac{S^2}{2}V(L\tau)+\frac{S^4}{8}\right) \\
 & = & K(L\tau,L\tau)-V(L\tau)^2-2\frac{m-1}{m}S^2V(L\tau)+\frac{3m-4}{m}\frac{S^4}{4}.
\end{array}
\end{equation}
Using Proposition~\ref{pro:momentV1}, we also find that the diagonal terms of $\Sigma^{Y}$ are 
\begin{equation}\label{eq:thm_S1_sigma11}
\Sigma^{Y}_{11}=\frac{1}{m}\left(K(L\tau,L\tau)-V(L\tau)^2\right) 
\end{equation}
and
\begin{equation}\label{eq:thm_S1_sigma22}
\Sigma^{Y}_{22}=K((M-1)\tau,(M-1)\tau)-V((M-1)\tau)^2.
\end{equation}
We can then apply the multivariate central limit theorem to the average vector defined in equation~\eqref{eq:thm_S1_vectSomme} and whose expected value is provided by Proposition~\ref{pro:momentV1}:
$$\sqrt{\frac{n}{M}}\left(\left(\begin{array}{c}
\widehat V_3(n,L) \\
\widehat V_3(n,M-1)
\end{array}\right) - \left(\begin{array}{c}
V(L\tau) \\
V((M-1)\tau)
\end{array}\right)\right)\overset{d}{\longrightarrow} \mathcal N_2\left(0,\Sigma^{Y}\right).$$
Then, we apply the delta method in the same way as in the general case and we have:
$$\sqrt{\frac{n}{M}}\left(\widehat{S^2_{1,3}}(n,L,M-1) - S^2\right)\overset{d}{\longrightarrow} \mathcal N\left(0,\frac{\gamma_{1,3}(L,M-1)}{M}\right),$$
with
$$\begin{array}{ccl}
\frac{\gamma_{1,3}(L,M-1)}{M} & = & \frac{4}{(M-1-L)^2}\left(\begin{array}{c}
M-1 \\
-L
\end{array}\right)^T\Sigma^{Y}\left(\begin{array}{c}
M-1 \\
-L
\end{array}\right) \\
 & = & \left\{\frac{(M-1)^2}{m}\left(K(L\tau,L\tau)-V(L\tau)^2\right) + L^2\left(K((M-1)\tau,(M-1)\tau)-V((M-1)\tau)^2\right)\right. \\
 & & \left. - 2(M-1)L\left(K(L\tau,L\tau)-V(L\tau)^2-2\frac{m-1}{m}S^2V(L\tau)+\frac{3m-4}{m}\frac{S^4}{4}\right)\right\} \frac{4}{(M-1-L)^2} \\
 & = & \frac{1}{M}\Gamma(L,M-1),
\end{array}$$
where we used equations~\eqref{eq:thm_S1_sigma12}, \eqref{eq:thm_S1_sigma11}, and~\eqref{eq:thm_S1_sigma22}. This proves the last result of the theorem.
\end{proof}

\section{Proofs for the market model with autocorrelated price increments}

\subsection{Proof of Proposition~\ref{pro:covIncrSq2}}\label{sec:covIncrSq2}

\begin{proof}
Like in the proof of Proposition~\ref{pro:covIncrSq1}, $K(u,\delta)16/S^4$ is equal to the quantity $\mathcal K^+$ when $\delta> 0$ or $\mathcal K^-$ when $\delta\leq 0$, as displayed in Lemma~\ref{lem:cokurtosis}, but with parameters adapted to Assumption~\ref{assum:fBm}: $\sigma_G^2=\frac{4u^{2H}\sigma^2}{S^2}$ and $\rho_G=c(\delta/u)$.

Like in the proof of Proposition~\ref{pro:covIncrSq1}, we study four possible cases, depending on the value of $\delta$. We start with the case $\delta< 0$, corresponding to strictly disjoint increments and thus, for all $i\neq j$, to $\rho_{i,j}=0$. The corresponding cokurtosis of Lemma~\ref{lem:cokurtosis} is $\mathcal K^-=(1+2\rho_G^2)\sigma_G^4 + 4\sigma_G^2 + 4$ so that
$$\begin{array}{ccl}
K(u,\delta) & = & (1+2\rho_G^2)u^{4H}\sigma^4+ u^{2H}\sigma^2S^2 + S^4/4 \\
 & = & V(u)^2 + 2\rho_G^2u^{4H}\sigma^4.
\end{array}$$ 

When $\delta=0$, the only difference with the previous case is $\rho_{3,2}=1$. This difference leads to $\mathcal K^-=(1+2\rho_G^2)\sigma_G^4 + 4(1-\rho_G)\sigma_G^2 + 4$ and thus 
$$\begin{array}{ccl}
K(u,\delta) & = & (1+2\rho_G^2)u^{4H}\sigma^4+ (1-\rho_G)u^{2H}\sigma^2S^2 + S^4/4 \\
 & = & V(u)^2 + 2\rho_G^2u^{4H}\sigma^4 -\rho_Gu^{2H}\sigma^2S^2.
\end{array}$$

When $\delta\in(0,u)$, we have, for all $i\neq j$, $\rho_{i,j}=0$ and we use the cokurtosis $\mathcal K^+=(1+2\rho_G^2)\sigma_G^4 + 4\sigma_G^2 + 4$ of Lemma~\ref{lem:cokurtosis}. We thus get the same result as in the case $\delta< 0$: $K(u,\delta) = V(u)^2+2\rho_G^2u^{4H}\sigma^4$.

Finally, when $\delta=u$, that is when the increments are the same, we have $\rho_G=1$, $\rho_{2,1}=\rho_{4,3}=1$, and $\rho_{i,j}=0$ for other values of $(i,j)$ with $i\neq j$. Then, $\mathcal K^+=3\sigma_G^4 + 12\sigma_G^2+8$ and
$$\begin{array}{ccl}
K(u,\delta) & = & 3u^{4H}\sigma^4+ 3u^{2H}\sigma^2S^2 + S^4/2 \\
 & = & V(u)^2+2u^{4H}\sigma^4+2u^{2H}\sigma^2S^2+S^4/4.
\end{array}$$ 
\end{proof}

\subsection{Proof of Proposition~\ref{pro:momentV2}}\label{sec:momentV2}

\begin{proof}
Like in the proof of Proposition~\ref{pro:momentV1}, the result on the first moment is immediate.

Regarding the second moment, we start with the case $v=2$. Following equation~\eqref{eq:momentV1_preuve1b} together with Proposition~\ref{pro:covIncrSq2}, we find
$$\begin{array}{ccl}
\E\left[\widehat V_2(n,L)^2\right] & = & \frac{1}{k_2(n,L)^2}\sum_{i=0}^{k_2(n,L)-1}\left\{ V(L\tau)^2+2(L\tau)^{4H}\sigma^4+2(L\tau)^{2H}\sigma^2S^2+S^4/4 \right. \\
 & & \left.+ 2\sum_{j=i+1}^{k_2(n,L)-1}\left(V(L\tau)^2 + 2c(i-j+1)^2(L\tau)^{4H}\sigma^4\right) -2c(0)(L\tau)^{2H}\sigma^2S^2 \right\} \\
 & = & V(L\tau)^2 + \frac{2(L\tau)^{4H}\sigma^4+2(1-c(0))(L\tau)^{2H}\sigma^2S^2+S^4/4}{k_2(n,L)} \\
 & & + \frac{4(L\tau)^{4H}\sigma^4}{k_2(n,L)^2}\sum_{i=0}^{k_2(n,L)-2} c(-i)^2(k_2(n,L)-1-i).
\end{array}$$
In addition, we can bound the following positive quantity
$$\frac{1}{k^2}\sum_{i=0}^{k-2} c(-i)^2(k-1-i) \leq \frac{1}{k}\sum_{i=0}^{k-2} c(-i)^2,$$
which, noting that $c(-k)=\mathcal O(k^{2H-2})$ when $k\rightarrow \infty$~\cite[Lemma 1]{Coeurjolly2001}, is $\mathcal O(k^{4H-3}/k)=o(1/k)$ if $H<3/4$, $\mathcal O(\log(k)/k)$ if $H=3/4$, and $\mathcal O(k^{4H-4})$ if $H>3/4$. This leads to the result displayed in Proposition~\ref{pro:momentV2} for $v=2$.

For $v=3$, using equation~\eqref{eq:momentV1_preuve1}, we only have to modify the values at which the function $c$ is considered, compared to the case $v=2$, in particular, $c(0)$ disappears because increments are strictly disjoint:
$$\begin{array}{ccl}
\E\left[\widehat V_3(n,L)^2\right] & = & V(L\tau)^2 + \frac{2(L\tau)^{4H}\sigma^4+2(L\tau)^{2H}\sigma^2S^2+S^4/4}{k_3(n,L)} \\
 & & + \frac{4(L\tau)^{4H}\sigma^4}{k_3(n,L)^2}\sum_{i=0}^{k_3(n,L)-2} c\left(-i-\frac{i+1}{L}\right)^2(k_3(n,L)-1-i).
\end{array}$$
Noting that $|c(-i-(i+1)/L)|<|c(-i)|$ for $i\geq 0$, we can use the same bound as in the case $v=2$ and we thus obtain the result displayed in Proposition~\ref{pro:momentV2}.

For $v=1$, since $K(u,\delta)$ is not a constant in $\delta$ when $\delta\leq 0$, we have to modify the decomposition of equation~\eqref{eq:momentV1_preuve1c} to make a distinction between non-overlapping increments: we thus incorporate $2(k_1(n,L)-L-l$ pairs of increments  of the type $(p_{u\tau}-p_0,p_{2u\tau+l\tau}-p_{u\tau+l\tau})$, with $l\geq 0$. Therefore
$$\begin{array}{ccl}
\E\left[\widehat V_1(n,L)^2\right] & = & \frac{1}{k_1(n,L)^2}\left(k_1(n,L)K(L\tau,L\tau) + 2\sum_{l=1}^{L-1}(k_1(n,L)-l)K(L\tau,(L-l)\tau) \right. \\
 & & \left.+ 2\sum_{l=0}^{k_1(n,L)-L}(k_1(n,L)-L-l)K(L\tau,-l\tau)\right) \\
 & = & V(L\tau)^2 + \frac{2(L\tau)^{4H}\sigma^4+2(L\tau)^{2H}\sigma^2S^2+S^4/4-(1-L/k_1(n,L))2c(0)(L\tau)^{2H}\sigma^2S^2}{k_1(n,L)} \\
 & & + \frac{4(L\tau)^{4H}\sigma^4}{k_1(n,L)^2}\sum_{1-L}^{k_1(n,L)-L} (k_1(n,L)-L-l)c(-l)^2. 
\end{array}$$
We can write the same argument as in the cases $v\in\{2,3\}$ to bound the sum of the $c(-l)^2$ and finally, when $H\in(0,3/4)$, we can write
$$\Var\left[\widehat V_1(n,L)\right]=\frac{2(L\tau)^{4H}\sigma^4-(2L\tau)^{2H}\sigma^2S^2+S^4/4}{k_1(n,L)} + o\left(\frac{1}{k_1(n,L)}\right),$$
thus leading to Proposition~\ref{pro:momentV2}.
\end{proof}

\subsection{Proof of Theorem~\ref{thm:S2}}\label{sec:S2}

\begin{proof}
The absence of bias in the case where $H$ is known is a direct consequence of Proposition~\ref{pro:momentV2} and of the fact that, according to equation~\eqref{eq:varianceIncrPrix_fBm}, $S^2=2(L'^{2H}V(L\tau)-L^{2H}V(L'\tau))/(L'^{2H}-L^{2H})$.

We prove the consistency of $\widehat{S^2_{2,v}}(n,L,L',H)$ and $\widehat{S^2_{2,v}}(n,L,L',\widehat H_{L''})$ for $v\in\{1,2,3\}$ in the same way as we did for Theorem~\ref{thm:S1}: following Proposition~\ref{pro:momentV2}, $\widehat V_v(n,L)$ converges in probability toward $V(L\tau)$ because its asymptotic quadratic risk is zero; then, the continuous mapping theorem concludes. 

Regarding the convergence in distribution of $\widehat{S^2_{2,v}}(n,L,L',H)$ or $\widehat{S^2_{2,v}}(n,L,L',\widehat H_{L''})$, we first establish it for $\widehat V_v(n,L)$. Like in the proof of Theorem~\ref{thm:S1}, and noting $\Delta$ the operator transforming the process in an increment of duration $L\tau$, the variables $(\Delta \eta_{i\tau})_i$ form an $\alpha$-mixing sequence, with $\alpha_n=0$ for $n$ sufficiently large, because of Assumption~\ref{assum:fBm}. The sequence $(\Delta p^{\star}_{i\tau})_i$, also has an $\alpha$-mixing property, which can be proved in this Gaussian case by the equivalent maximal correlation mixing, with an $\alpha$ dominated by the correlation~\cite[Theorem 2]{KR}. The variables $\Delta p^{\star}_{i\tau}$ and $\Delta \eta_{i\tau}$ being independent to each other (Assumption~\ref{assum:indepProcesses}) and mixing, $(\Delta p^{\star}_{i\tau},\Delta \eta_{i\tau})_i$ and thus $((\Delta p^{\star}_{i\tau}+\Delta \eta_{i\tau})^2)_i$ are mixing sequences with a parameter $\alpha$ lower than the sum of the $\alpha$ of each of the two components~\cite[Theorem 5.1]{Bradley}. In other words, since $\alpha_n=0$ for $\Delta \eta_{i\tau}$ and $n$ sufficiently large, the mixing condition of $(\Delta p^{\star}_{i\tau}+\Delta \eta_{i\tau})^2$ is the same as the one of $\Delta p^{\star}_{i\tau}$ or of $(\Delta p^{\star}_{i\tau})^2$. As a consequence, a central limit theorem holds for $\widehat V_v(n,L)$, whatever $v$, exactly in the same conditions as if the $\eta_{i\tau}$ were replaced by 0, that is as soon as $H<3/4$~\cite[Proposition 1]{Coeurjolly2001}. This asymptotic property on $\widehat V_v(n,L)$ also holds for vectors corresponding to various lags, such as $(\widehat V_v(n,L),\widehat V_v(n,L'),\widehat V_v(n,L''),\widehat V_v(n,2L''),\widehat V_v(n,4L''))^T$~\cite[Proposition 3]{Coeurjolly2001}.

Finally, we conclude about the convergence in distribution of $\widehat{S^2_{2,v}}(n,L,L',H)$ and $\widehat{S^2_{2,v}}(n,L,L',\widehat H_{L''})$ using the delta method. In particular, when $H$ is known, the framework is very similar to the one developed in the proof of Theorem~\ref{thm:S1}, except that we consider the function  $g:(x,y)\mapsto 2(L'^{2H}x-L^{2H}y)/(L'^{2H}-L^{2H})$, whose gradient is
$$\nabla g(x,y)=\frac{2}{L'^{2H}-L^{2H}}\left(\begin{array}{c}
L'^{2H} \\
-L^{2H}
\end{array}\right).$$
Finally, we get
$$\sqrt{n}\left(\widehat{S^2_{2,v}}(n,L,L',H) - S^2\right)\overset{d}{\longrightarrow} \mathcal N\left(0,\gamma_{2,v}(L,L',H)\right),$$
with
$$\begin{array}{ccl}
\gamma_{2,v}(L,L',H) & = & \frac{4}{(L'^{2H}-L^{2H})^2} \Bigl[L'^{4H}\zeta_v(L)\sigma_{2,v}^2(L) + L^{4H}\zeta_v(L')\sigma_{2,v}^2(L') \\
 & & - 2(LL')^{2H}\sqrt{\zeta_v(L)\zeta_v(L')}\sigma_{2,v}(L)\sigma_{2,v}(L')r_{2,v}(L,L')\Bigr].
\end{array}$$

In the case where $H$ is unknown, we focus on $(L,L',L'')=(1,2,1)$. Let $h(x,y,z)=\log_2((z-y)/(y-x))$. In equation~\eqref{eq:estimH}, we have added an absolute value, which will not be required asymptotically for the delta method, because the limit of $(\widehat V_v(n,4L)-\widehat V_v(n,2L))/(\widehat V_v(n,2L)-\widehat V_v(n,L))$ is strictly positive. We first remark that $x^{\log y}=y^{\log x}$, so that $2^{h(x,y,z)}=(z-y)/(y-x)$. The delta method will be applied to the mapping
$$\begin{array}{ccl}
g(x,y,z) & = & 2\left(2^{h(x,y,z)}x-1^{h(x,y,z)}y\right)/\left(2^{h(x,y,z)}-1^{h(x,y,z)}\right) \\
 & = & 2(zx-y^2)/(z-2y+x),
\end{array}$$
whose gradient is
$$\nabla g(x,y,z)=\frac{2}{(z-2y+x)^2}\left(\begin{array}{c}
(z-y)^2 \\
2y(y-z-x)+2zx \\
(x-y)^2
\end{array}\right),$$
and to the vector $(\widehat V_v(n,1),\widehat V_v(n,2),\widehat V_v(n,4))^T$, which is such that
\begin{equation}\label{eq:TCL_dim3}
\sqrt{n}\left(\left(\begin{array}{c}
\widehat V_v(n,1) \\
\widehat V_v(n,2) \\
\widehat V_v(n,4)
\end{array}\right) - \left(\begin{array}{c}
V(\tau) \\
V(2\tau) \\
V(4\tau)
\end{array}\right)\right)\overset{d}{\longrightarrow} \mathcal N_3\left(0,\Sigma\right),
\end{equation}
where $\mathcal N_3$ is the trivariate Gaussian distribution. The delta method directly leads to equation~\eqref{eq:TCL_fBm}.
\end{proof}

\section{Proofs for the market model with autocorrelated trades}

\subsection{Proof of Proposition~\ref{pro:correlOU}}\label{sec:correlOU}

\begin{proof}
The solution of equation~\eqref{eq:OrnsteinSDE} is
$$\eta_t^{\star}=\frac{\Xi}{\sqrt{2\theta}}e^{-\theta t}W_{e^{2\theta t}}.$$
Noting that $\E[\widetilde\eta_t]=0$ and $\Var[\widetilde\eta_t]=S^2$, the autocorrelation of the discrete-valued process is
\begin{equation}\label{eq:proofCovMicro}
\begin{array}{ccl}
\corr\left(\widetilde\eta_s,\widetilde\eta_t\right) & = & \E[\sgn(\eta_s^{\star}\eta_t^{\star})] \\
 & = & \proba\left(W_{e^{2\theta t}}W_{e^{2\theta s}} > 0\right) - \proba\left(W_{e^{2\theta t}}W_{e^{2\theta s}} < 0\right) \\
 & = & 1-2\proba\left(W_{e^{2\theta t}}W_{e^{2\theta s}} < 0\right),
\end{array}
\end{equation}
by complementarity. For $x>0$ and $0<u<v$, we have
$$\begin{array}{ccl}
\proba(W_uW_v<0|W_u=x) & = & \proba\left(\left.\frac{W_v-W_u}{\sqrt{v-u}}<\frac{-x}{\sqrt{v-u}}\right|W_u=x\right) \\
 & = & N\left(\frac{-x}{\sqrt{v-u}}\right),
\end{array}$$
where $N$ is the standard Gaussian cdf. Now, if $x<0$, we get
$$\begin{array}{ccl}
\proba(W_uW_v<0|W_u=x) & = & \proba\left(\left.\frac{W_v-W_u}{\sqrt{v-u}}>\frac{-x}{\sqrt{v-u}}\right|W_u=x\right) \\
 & = & N\left(\frac{x}{\sqrt{v-u}}\right).
\end{array}$$
By the law of total probability, we can thus write the following, noting $\alpha=((v/u)-1)^{-1/2}$, $g_{u}$ the Gaussian pdf of variance u, and applying a change of variable $y=-x/\sqrt{u}$ in the first integral and $y=x/\sqrt{u}$ in the second:
$$\begin{array}{ccl}
\proba(W_uW_v<0) & = & \int_{-\infty}^{0}N(\alpha x/\sqrt{u})g_u(x)dx + \int_{0}^{+\infty}N(-\alpha x/\sqrt{u})g_u(x)dx \\
 & = & 2\int_{0}^{+\infty}N(-\alpha y)g_1(y)dy \\
 & = & \frac{1}{2}+\frac{1}{\pi}\arctan(-\alpha),
\end{array}$$
where we prove the last line with Lemma 1 in~\cite{Garcin22}. Using equation~\eqref{eq:proofCovMicro} and replacing $u$ and $v$ by $e^{2\theta s}$ and $e^{2\theta t}$, we thus conclude that 
$$\corr\left(\widetilde\eta_s,\widetilde\eta_t\right)=\frac{2}{\pi}\arctan\left(\frac{1}{\sqrt{e^{2\theta (t-s)}-1}}\right).$$
\end{proof}

\subsection{Proof of Proposition~\ref{pro:covIncrSq3}}\label{sec:covIncrSq3}

\begin{proof}
Like in the proof of Propositions~\ref{pro:covIncrSq1} and~\ref{pro:covIncrSq2}, $K(u,\delta)16/S^4$ is equal to the quantity $\mathcal K^+$ when $\delta> 0$ or $\mathcal K^-$ when $\delta\leq 0$, as displayed in Lemma~\ref{lem:cokurtosis}, with parameters adapted to Assumption~\ref{assum:correlTrades}: $\sigma_G^2=\frac{4u\sigma^2}{S^2}$ and correlations depending on $\delta$. 

When $\delta\in(0,u]$, we have overlapping increments and $\rho_G=\delta/u$, $\rho_{2,1}=\rho_{4,3}=e^{-(u-\delta)/\lambda}$, $\rho_{3,1}=\rho_{4,2}=e^{-u/\lambda}$, $\rho_{3,2}=e^{-\delta/\lambda}$, and $\rho_{4,1}=e^{-(2u-\delta)/\lambda}$. In this case, the relevant cokurtosis in Lemma~\ref{lem:cokurtosis} is 
$$\begin{array}{ccl}
\mathcal K^+ & = & \left(1+2\frac{\delta^2}{u^2}\right)\sigma_G^4 + 4\sigma_G^2 \left(1-e^{-u/\lambda}\right) \\
 & & +  4\left(1-2e^{-u/\lambda}+e^{-2(u-\delta)/\lambda}\right) + 4\frac{\delta}{u}\sigma_G^2\left(2e^{-(u-\delta)/\lambda}-e^{-\delta/\lambda}-e^{-(2u-\delta)/\lambda}\right)
\end{array}$$
and we finally get
$$\begin{array}{ccl}
K(u,\delta) & = & \left(1+2\frac{\delta^2}{u^2}\right)u^2\sigma^4+ u\sigma^2S^2 \left(1-e^{-u/\lambda}\right) \\
 & & + \frac{S^4}{4}\left(1-2e^{-u/\lambda}+e^{-2(u-\delta)/\lambda}\right) + \delta \sigma^2S^2\left(2e^{-(u-\delta)/\lambda}-e^{-\delta/\lambda}-e^{-(2u-\delta)/\lambda}\right) \\
 & = & V(u)^2 + 2\delta^2\sigma^4 + \frac{S^4}{4}\left(e^{-2(u-\delta)/\lambda}-e^{-2u/\lambda}\right) + \delta \sigma^2S^2\left(2e^{-(u-\delta)/\lambda}-e^{-\delta/\lambda}-e^{-(2u-\delta)/\lambda}\right).
\end{array}$$ 

In the case of non-overlapping increments, that is for $\delta\leq 0$, we have $\rho_G=0$, $\rho_{2,1}=\rho_{4,3}=e^{-u/\lambda}$, $\rho_{3,1}=\rho_{4,2}=e^{-(u-\delta)/\lambda}$, $\rho_{3,2}=e^{\delta/\lambda}$, and $\rho_{4,1}=e^{-(2u-\delta)/\lambda}$. In this case, the relevant cokurtosis in Lemma~\ref{lem:cokurtosis} is 
$\mathcal K^- = \sigma_G^4+4\sigma_G^2\left(1-e^{-u/\lambda}\right) + 4\left(1-e^{-u/\lambda}\right)^2$, which more simply writes $\left(\sigma_G^2+2(1-e^{-u/\lambda})\right)^2$. We thus get $K(u,\delta)=V(u)^2$.
\end{proof}

\subsection{Proof of Proposition~\ref{pro:momentV3}}\label{sec:momentV3}

\begin{proof}
Like in the proof of Propositions~\ref{pro:momentV1} and~\ref{pro:momentV2}, the result on the first moment is immediate.

For $v=3$ and $v=2$, using respectively equation~\eqref{eq:momentV1_preuve1} and equation~\eqref{eq:momentV1_preuve1b}, along with Proposition~\ref{pro:covIncrSq3}, the terms $K(L\tau,(i-j+1)L\tau)$ or $K(L\tau,(i-j+(i-j+1))L\tau)$ being all equal to $K(L\tau,0)$, we get
$$\Var\left[\widehat V_v(n,L)\right]=\frac{2\delta^2\sigma^4 + \frac{S^4}{4}\left(e^{-2(u-\delta)/\lambda}-e^{-2u/\lambda}\right)
+ \delta \sigma^2S^2\left(2e^{-(u-\delta)/\lambda}-e^{-\delta/\lambda}-e^{-(2u-\delta)/\lambda}\right)}{k_v(n,L)},$$
with $u=\delta=L\tau$, thus leading to
$$\Var\left[\widehat V_v(n,L)\right]=\frac{2L^2\tau^2\sigma^4 + \frac{S^4}{4}\left(1-e^{-2L\tau/\lambda}\right)
+ 2L\tau \sigma^2S^2\left(1-e^{-L\tau/\lambda}\right)}{k_v(n,L)},$$

For $v=1$, since $K(u,\delta)=K(u,0)$ for $\delta\leq 0$, we can use the same decomposition as in equation~\eqref{eq:momentV1_preuve1c}:
$$\begin{array}{ccl}
\E\left[\widehat V_1(n,L)^2\right] & = & \frac{1}{k_1(n,L)^2}\left(k_1(n,L)K(L\tau,L\tau) + \sum_{m=1}^{L-1}2(k_1(n,L)-L+m)K(L\tau,m\tau) \right. \\
 & & \left.+ \left(k_1(n,L)^2-k_1(n,L)-\sum_{l=1}^{L-1}2(k_1(n,L)-l)\right)K(L\tau,0)\right) \\
 & = & V(L\tau)^2 + \frac{1}{k_1(n,L)^2}\left(k_1(n,L)\left(2L^2\tau^2\sigma^4 + \frac{S^4}{4}\left(1-e^{-2L\tau/\lambda}\right)
+ 2L\tau \sigma^2S^2\left(1-e^{-L\tau/\lambda}\right)\right) \right. \\
 & & \left. + 2\sum_{m=1}^{L-1}(k_1(n,L)-L+m) \left[2m^2\tau^2\sigma^4 + \frac{S^4}{4}e^{-2L\tau/\lambda}\left(e^{2m\tau/\lambda}-1\right) \right.\right. \\
 & & \left.\left.+ m\tau \sigma^2S^2\left(2e^{-(L-m)\tau/\lambda}-e^{-m\tau/\lambda}-e^{-(2L-m)\tau/\lambda}\right) \right]\right) 
\end{array}$$
Subtracting $V(L\tau)^2$ and noting that $\sum_{m=1}^{L-1}e^{xm}=f_L(x)$ and $\sum_{m=1}^{L-1}me^{xm}=f_L'(x)$, we get the variance of $\widehat V_1(n,L)$:
$$\begin{array}{ccl}
\Var\left[\widehat V_1(n,L)\right] & = & \frac{1}{k_1(n,L)}\left(2L^2\tau^2\sigma^4 + \frac{S^4}{4}\left(1-e^{-2L\tau/\lambda}\right) + 2L\tau \sigma^2S^2\left(1-e^{-L\tau/\lambda}\right) \right. \\
 & & \left. + 4\tau^2\sigma^4\sum_{m=1}^{L-1} m^2 + \frac{S^4}{2}e^{-2L\tau/\lambda}\sum_{m=1}^{L-1}\left(e^{2m\tau/\lambda}-1\right)  \right. \\
 & & \left. + 2\tau \sigma^2S^2\sum_{m=1}^{L-1}m\left(2e^{-(L-m)\tau/\lambda}-e^{-m\tau/\lambda}-e^{-(2L-m)\tau/\lambda}\right) \right) + \mathcal{O}\left(\frac{1}{k_1(n,L)^2}\right) \\
  & = & \frac{1}{k_1(n,L)}\left(2L^2\tau^2\sigma^4 + \frac{S^4}{4}\left(1-e^{-2L\tau/\lambda}\right) + 2L\tau \sigma^2S^2\left(1-e^{-L\tau/\lambda}\right) \right. \\
 & & \left. + \frac{2}{3}\tau^2\sigma^4(L-1)L(2L-1) + \frac{S^4}{2}e^{-2L\tau/\lambda}\left(f_L\left(\frac{2\tau}{\lambda}\right)-L+1\right)  \right. \\
 & & \left. + 2\tau \sigma^2S^2\left(2e^{-L\tau/\lambda}f_L'\left(\frac{\tau}{\lambda}\right) - f_L'\left(-\frac{\tau}{\lambda}\right) -e^{-2L\tau/\lambda}f_L'\left(\frac{\tau}{\lambda}\right)\right) \right) + \mathcal{O}\left(\frac{1}{k_1(n,L)^2}\right).
\end{array}$$
Noting that $k_1(n,L)=n-L$, we get the expected result.
\end{proof}

\subsection{Proof of Theorem~\ref{thm:S3}}\label{sec:S3}

\begin{proof}
The absence of bias in the case where $\lambda$ is known is a direct consequence of Proposition~\ref{pro:momentV3} and of equation~\eqref{eq:varianceIncrPrix_correlTrades}.

We prove the consistency of $\widehat{S^2_{3,v}}(n,L,L',\rho)$ and $\widehat{S^2_{3,v}}(n,L,L',(\widehat{\rho^{L''}})^{1/L''})$ for $v\in\{1,2,3\}$ in the same way as we did for Theorem~\ref{thm:S1}: following Proposition~\ref{pro:momentV3}, $\widehat V_v(n,L)$ converges in probability toward $V(L\tau)$ because its asymptotic quadratic risk is zero; then, the continuous mapping theorem concludes.

Regarding the convergence in distribution, since the sequence of observed prices is $\alpha$-mixing, with $\alpha_n$ exponentially decaying, we can apply the central limit theorem~\cite[Theorem 27.4]{Billingsley}, with an expression similar to equation~\eqref{eq:TCL_dim2} for the vector $(\widehat V_v(n,L),\widehat V_v(n,L'))^T$. Our estimator of $S^2$, in the case where $\rho$ is known, is obtained with a function 
$$g:(x,y)\mapsto \frac{2(L' x-L y)}{L'(1-\rho^L)-L(1-\rho^{L'})}$$
applied to $(\widehat V_v(n,L),\widehat V_v(n,L'))$. This function $g$ is the same as the one introduced in the proof of Theorem~\ref{thm:S1}, up to the factor $(L'-L)/(L'(1-\rho^L)-L(1-\rho^{L'}))$. Therefore, obtaining $\gamma_{3,v}(L,L',\rho)$ from the expression of $\gamma_{1,v}(L,L')$ is straightforward. This proves equation~\eqref{eq:TCL3}.

When $\rho>0$ is unknown, we can write, for $n$ large enough, the estimator of equation~\eqref{eq:S3_correlUnknown} as $\widehat{S^2_{3,v}}(n,L)=h(\widehat V_v(n,L),\widehat V_v(n,2L),\widehat V_v(n,4L))$, where $h$ is defined in equation~\eqref{eq:deltafunction3} and omit the absolute values introduced in equation~\eqref{eq:S3_correlUnknown}. Once again, starting from the multivariate central limit theorem of the vector $(\widehat V_v(n,L),\widehat V_v(n,2L),\widehat V_v(n,4L))^T$, as in equation~\eqref{eq:TCL_dim3}, we can apply the delta method for the function $h$, leading to the result displayed in Theorem~\ref{thm:S3}.
\end{proof}

\end{document}